\documentclass{aastex63}

\usepackage{CJK}
\usepackage{rotating}
\usepackage{multirow}
\usepackage{amsmath,bm}

\submitjournal{ApJ}

\shorttitle{}
\shortauthors{Su et al.}

\begin{document}
\begin{CJK*}{UTF8}{gbsn}
	
\title{Quantifying the Magnetic Structure of a Coronal Shock Producing a Type II Radio Burst  }

\correspondingauthor{Su, W.}
\email{suwei25@mail.sysu.edu.cn}

\author{W. Su (苏威)}
\affiliation{TianQin Research Center for Gravitational Physics \& School of Physics and Astronomy, Sun Yat-sen University (Zhuhai Campus), Zhuhai 519082, P.R. China}
\affiliation{MOE Key Laboratory of Fundamental Physical Quantities Measurements, Hubei Key Laboratory of Gravitation and Quantum Physics, PGMF and School of Physics, Huazhong University of Science and Technology, Wuhan 430074, China}
\affiliation{Key Laboratory for Modern Astronomy and Astrophysics (Nanjing University), Ministry of Education, Nanjing 210023, China}

\author{T. M. Li (李汤姆)}
\affiliation{Key Laboratory for Modern Astronomy and Astrophysics (Nanjing University), Ministry of Education, Nanjing 210023, China}

\author{X. Cheng (程鑫)}
\affiliation{Key Laboratory for Modern Astronomy and Astrophysics (Nanjing University), Ministry of Education, Nanjing 210023, China}

\author{L. Feng (封莉)}
\affiliation{Key Laboratory of Dark Matter and Space Astronomy, Purple Mountain Observatory, Nanjing 210033, China}

\author{P. J. Zhang (张沛锦)}
\affiliation{CAS Key Laboratory of Geospace Environment, School of Earth and Space Sciences, University of Science and Technology of China (USTC), Hefei, Anhui 230026, People’s Republic of China}

\author{P. F. Chen (陈鹏飞)}
\author{M. D. Ding (丁明德)}
\affiliation{Key Laboratory for Modern Astronomy and Astrophysics (Nanjing University), Ministry of Education, Nanjing 210023, China}

\author{L. J. Chen (陈林杰)}
\affiliation{Key Laboratory of Solar Activity, National Astronomical Observatories of Chinese Academy of Sciences, Beijing 100012, China}

\author{Y. Guo (郭洋)}
\affiliation{Key Laboratory for Modern Astronomy and Astrophysics (Nanjing University), Ministry of Education, Nanjing 210023, China}

\author{Y. Wang (王炎)}
\affiliation{MOE Key Laboratory of Fundamental Physical Quantities Measurements, Hubei Key Laboratory of Gravitation and Quantum Physics, PGMF and School of Physics, Huazhong University of Science and Technology, Wuhan 430074, China}

\author{D. Li (李东)}
\affiliation{Key Laboratory of Dark Matter and Space Astronomy, Purple Mountain Observatory, Nanjing 210033, China}

\author{L. Y. Zhang (张蠡岳)}
\affiliation{Key Laboratory for Modern Astronomy and Astrophysics (Nanjing University), Ministry of Education, Nanjing 210023, China}








\begin{abstract}

Type II radio bursts are thought to be produced by shock waves in the solar atmosphere. However, what magnetic conditions are needed for the generation of type II radio bursts is still a puzzling issue. Here, we quantify the magnetic structure of a coronal shock associated with a type II radio burst. Based on the multi-perspective extreme-ultraviolet observations, we reconstruct the three-dimensional (3D) shock surface. By using a magnetic field extrapolation model, we then derive the orientation of the magnetic field relative to the normal of the shock front ($\theta_{\rm Bn}$) 
and Alfv\'{e}n Mach number ($M_A$) on the shock front. 
Combining the radio observations from Nancay Radio Heliograph, we obtain the source region of the type II radio burst on the shock front. It is found that the radio burst is generated by
a shock with $M_A \gtrsim 1.5$ and a bimodal distribution of $\theta_{Bn}$.
We also use the Rankine-Hugoniot relations to quantify the properties of the shock downstream. 
Our results provide a quantitative 3D magnetic structure condition of a coronal shock that produces a type II radio burst.

\end{abstract}



\section{Introduction} \label{sec:intro}

Magnetohydrodynamic (MHD) shocks are an effective accelerator for charged particles \citep{Draine1993,Hao2017}. 
A typical example is the coronal shock. Coronal shocks are able to generate energetic electron beams, and cause type II radio bursts \citep{Wild1950,Zheleznyakov1970,Du2014}. 
Several issues are related to this topic. The first one is how shock waves are generated. The second one is how electrons are accelerated in the shocks, and the third is how the radio emission is produced by the energetic electrons. 

The generation mechanism of coronal shocks is still under debate.
It was proposed to be generated by a piston wave driven by CMEs \citep{Chen2011,Ying2019} or a blast wave excited by the pressure pulse of flares \citep{Vrsnak2008,Magdalenic2012}. Besides, coronal shocks were also able to be generated by jets or magnetic reconnection outflows \citep{Su2015,ChenB2015}.
Regardless of what drives coronal shocks, the conditions of shocks that are able to generate type II radio bursts should be similar. 
How shock waves accelerate electrons is another issue under debate \citep{Masters2013}. It seems that the dominant electron acceleration mechanism is different for quasi-perpendicular and quasi-parallel shocks. For example, shock drift acceleration (SDA) was claimed to be efficient to accelerate electrons in quasi-perpendicular shocks \citep{Mann2018,Kong2020}, while the first-order Fermi acceleration plays a major role in quasi-parallel shocks \citep{Mann2001,Qin2018}. Correspondingly, type II radio bursts have been found to be generated by both quasi-perpendicular and quasi-parallel
shocks \citep{Mann1995b,Maguire2020}. However, it is still unclear which case is more prevailing. Once electrons are accelerated locally by shock waves, electromagnetic wave emissions would be generated by either plasma emission and electron cyclotron maser emission mechanisms, forming type II radio bursts \citep{Ginzburg1958,Wu1986,Zhao2014}.

Considering that type II radio bursts are often generated at certain regions around the shock front rather than over a wide region of the shock front \citep{Su2016,Lu2017,Zucca2018,Morosan2020}, it seems that the local parameters of a shock front can significantly affect the generation of radio bursts.
Alfv\'{e}n Mach number ($M_{A}$) is a physical parameter to describe the strength of MHD shocks. 
Qualitatively, type II radio bursts were reported to be likely generated at the regions where the local Alfv\'{e}n speed ($v_{A}$) is low \citep{Gopalswamy2009}, or to be likely generated when the `pistons', CMEs, are faster \citep{Gopalswamy2005,Lee2014}, or to be likely generated at places where the density is high \citep{Reiner2003}. The common feature among these scenarios is that $M_{A}$ can easily become large. 
Based on the differential emission measure (DEM) method \citep{Weber2004,SuY2018} and the magnetic field extrapolation model, a map of $v_{A}$ or shock Mach number in the corona can be obtained \citep{Zucca2014a,Rouillard2016}. It has been shown quantitatively that type II radio bursts are indeed found to be generated at the regions where $M_{A}$ is large \citep{Su2016}.

Since there are no in-situ observations in the source region of type II radio burst in the corona, it is difficult for us to obtain the magnetic field around the coronal shock and to identify whether the shock is quasi-perpendicular or quasi-parallal. In the low corona, coronal shocks are usually approximately to be quasi-perpendicular ($<1.5~\rm R_{\odot}$) \citep{Ma2011,Su2016}. However, in the field-of-view of the Large Angle Spectrometric Coronagraph onboard the Solar and Heliospheric Observatory \citep[SOHO/LASCO;][]{Brueckner1995}, the magnetic field in the shock upstream is usually approximated to be radial \citep{Bemporad2010,Susino2015}. 
Besides, the fine structures of type II radio bursts in the radio dynamic spectra can help us infer the orientation of magnetic field on the shock fronts \citep{Mann2005}.
Recently, magnetic field extrapolation models have been used in the study on type II radio bursts \citep{Zucca2014b,Zucca2018,Morosan2019}, this method can help reveal the magnetic conditions around the coronal shock that is associated with type II radio bursts.

In this paper, we use extreme-ultraviolet (EUV) observations of the Solar Dynamics Observatory \citep[SDO;][]{Pesnell2012} and the Solar Terrestrial Relations Observatory \citep[STEREO;][]{Kaiser2008} to construct a three-dimensional (3D) model of a coronal shock front, which is associated with a type II radio burst. Combined with the radio observations from Nancay Radio Heliograph \citep[NRH;][]{Kerdraon1997}, we determine the position of the source region of the type II radio burst. 
The Potential-Field-Source-Surface (PFSS) model \citep{Schatten1969,Wang1992,Schrijver2003} is used to reconstruct the magnetic structure of the type II radio burst. Finally, we derive the magnetic conditions for the source region of the type II radio burst.

\section{Observations} \label{sec:2}

\begin{figure}[ht]
	\centering
	\includegraphics[width=16 cm]{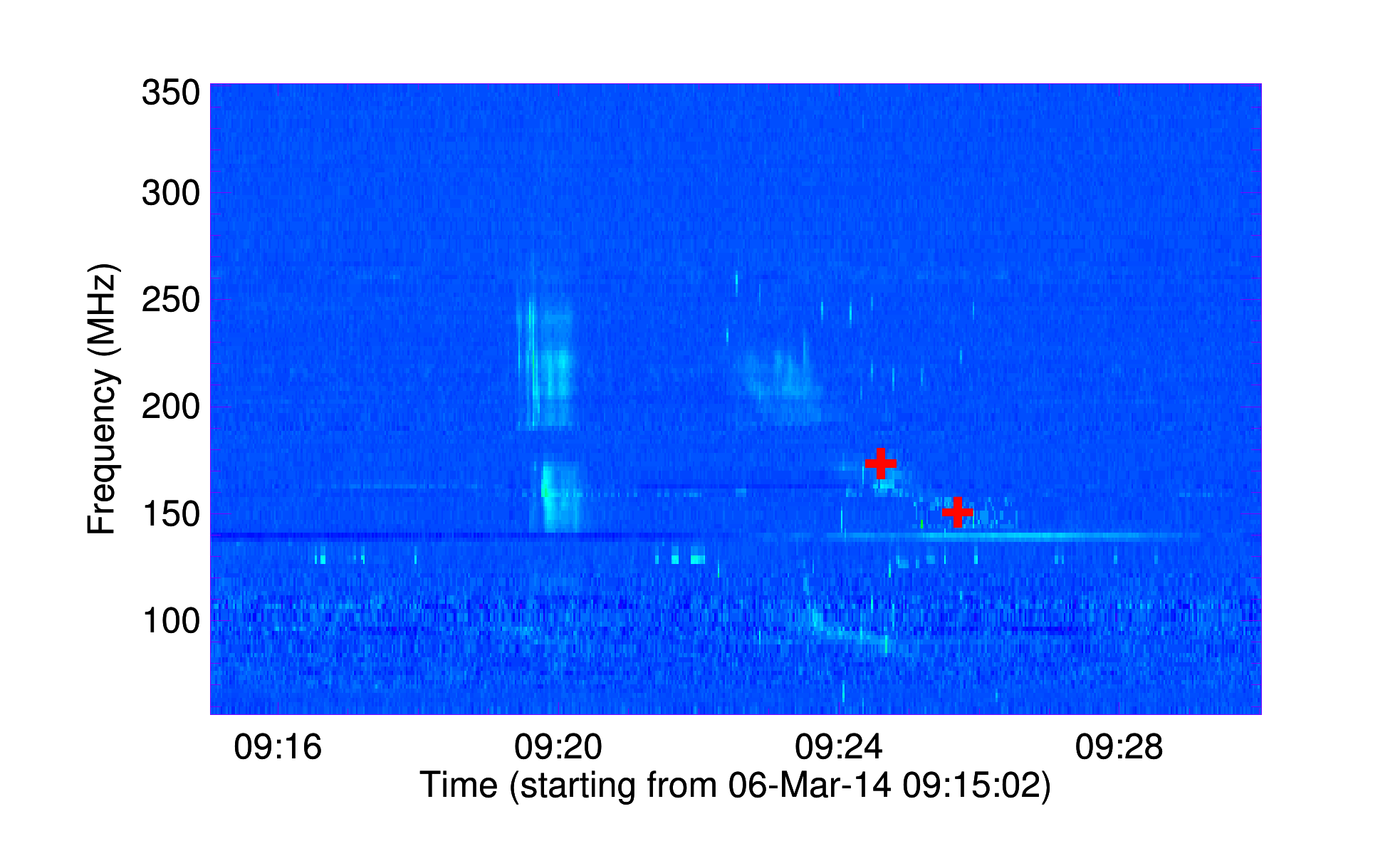}
	\caption{Dynamic spectrum of the type II radio burst on 2014 March 6. The frequencies corresponding to the red pluses are 173.2 MHz and 150.9 MHz on the lane of the type II radio burst from the NRH observations. The time corresponding to the red pluses are 09:24:42  and 09:26:18 UT from the AIA observations.}
	\label{fig:spec}
\end{figure}

The dynamic spectrum of the type II radio burst is shown in Figure \ref{fig:spec}. 
There are two slowly-drifting lanes in the dynamic spectrum, and the upper one (the harmonic component) has a frequency about twice that of the lower one (the fundamental component). The harmonic lane is more prominent than the fundamental lane, and has a start time of 09:23 UT on Mar 6, 2014, with a start frequency of 225 MHz. Below 145 MHz, the harmonic lane becomes too diffuse to be distinguished. 

\begin{figure}[ht]
	\centering
	\includegraphics[width=10 cm]{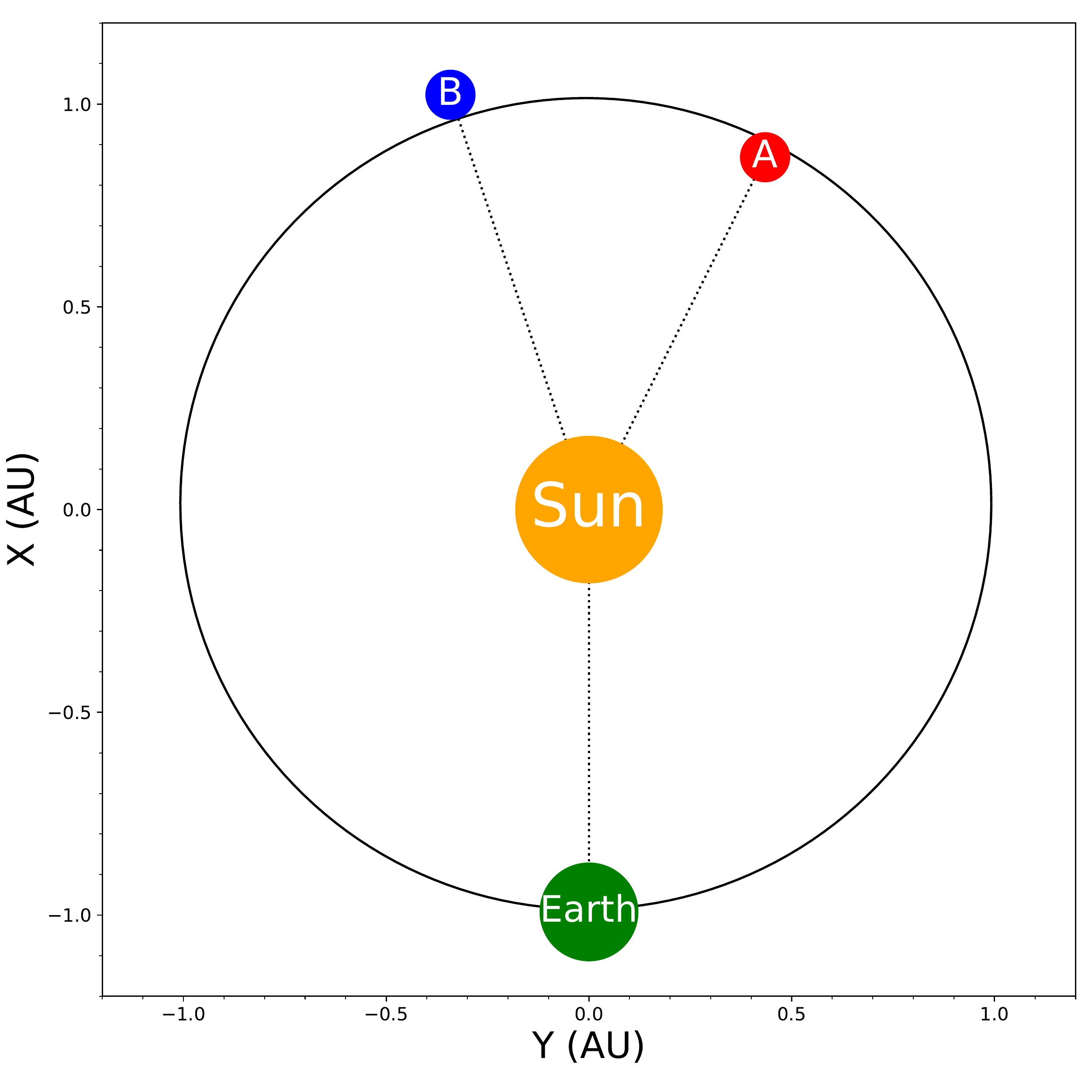}
	\caption{The schematic diagram of the positions of ST\_A and ST\_B on the ecliptic plane in the heliocentric coordinate system on 2014 March 06. The orange disk in the center represents the Sun, the green disk denotes the Earth, and the red and blue dots denote ST\_A and ST\_B, respectively.}
	\label{fig:pos}
\end{figure}

At the time of this event, the relative position of SDO, the STEREO $Ahead$ (ST\_A) and $Behind$ (ST\_B) on the ecliptic in the heliocentric coordinate system are shown in Figure \ref{fig:pos}, and the separation angle between the SDO and the ST\_A/ST\_B are 153$^\circ$ and 161$^\circ$, respectively.
The associated flare, starting from 09:21 UT, about 2 minutes before the type II radio burst, is located near the east limb of the solar disk in the field of view of the Atmospheric Imaging Assemble \citep[AIA;][]{Lemen2012} onboard the SDO, which is shown in Figure \ref{fig:EUV-TD}a. Therefore, the eruption can only be observed by SDO and ST\_B.  Figure \ref{fig:sh-fit2} shows the running difference images at 193 \AA~ of SDO/AIA and at 195 \AA~ of the Extreme Ultraviolet Imager \citep[EUVI;][]{Wuelser2004} onboard the ST\_B. 
Note that the 193 \AA\ images of AIA in Figures \ref{fig:sh-fit2}b and \ref{fig:sh-fit2}d
are made at the moments of 09:24:42 and 09:26:18 UT, which are annotated as red pluses in Figure \ref{fig:spec}.
The eruption is at the west limb in the field of view of ST\_B (Figures \ref{fig:sh-fit2}a and \ref{fig:sh-fit2}c). 
As indicated by the AIA 193 \AA~ images (Figures \ref{fig:EUV-TD}a, \ref{fig:sh-fit2}b and \ref{fig:sh-fit2}d), the leading edge of the CME is clearly visible.
In particular, there is a distinct feature in front of the CME leading edge. 
Such a feature has been identified to be the CME-driven shock in a number of studies \citep{Ma2011,Vourlidas2013,Lee2014,Su2016,Feng2020}. 
The shape of the CME-driven shock is quite regular, which is favourable for fitting the 3D shock structure. 

The observations of the NRH can provide radio maps at multiple bands (150.9, 173.2, 228.0, 270.6, 298.7, 327.0, 360.8, 408.0, 432.0, 445.0 MHz). 
We use the NRH package in the SolarSoftWare to produce calibrated radio images of the Sun, by which we obtain the radio images with a temporal resolution of 1 s.
The radio images of the NRH is discontinuous in frequency, thus we can only get the radio maps when the NRH observation bands lie in the range of the type II radio burst. 
The NRH radio maps at 173.2 and 150.9 MHz are used for the two moments in Figures \ref{fig:sh-fit2}b and \ref{fig:sh-fit2}d, which are also marked in Figure \ref{fig:spec} as red pluses. 
The radio maps at separate frequencies 173.2 MHz (09:24:42 UT) and 150.9 MHz (09:26:18 UT) are shown in Figure \ref{fig:nrh}. The purple, green and red contour lines correspond to 50\%, 80\% and 90\% of the brightness temperature maximum.

\section{Results}

\subsection{Fitting the shock surface}

Coronal shocks can be approximated to have an symmetric 3D bow-shock geometry \citep{Ontiveros2009,Chen2014}. In the cylindrical coordinate systems, the shape of the bow shock can be given by the following formula \citep{Smith2003}:
\begin{equation}
\label{equation1}
Z = h -  \frac{d}{s} \times (\frac{\sqrt{R^2}}{d})^s, \\
\end{equation}
where $h$ describes the apex height of the shock, $s$ determines the opening angle of the shock, and $d$ is the $semilatus~rectum$ that controls the width of the shock. 
The $Z$-axis in the heliocentric coordinate system is assumed to be the initial symmetrical axis. Besides, we input two sets of angular parameters to fit the shock shape, the latitude and longitude ($\theta, \phi$) of the eruption location with the corresponding base vector $\bm{e}$, and the latitude and longitude ($\theta', \phi'$) of the eruption direction with the corresponding base vector $\bm{e}'$. According to ($\theta, \phi$) and ($\theta', \phi'$), the rotation matrices can be constructed to rotate the shock to the correct position. 
Note that there is a problem known as $gimbal~lock$, which is the loss of one degree of freedom in the 3D space. It happens when two of the rotation axes are driven into a parallel configuration, and the system degenerates into rotating in the 2D space just like ``locking''.

In order to avoid the $gimbal~lock$ problem, we do not take the $X$-, $Y$-, and $Z$-axes in the coordinate system as the rotation axes. For example, when we rotate the fitting structure from the position ($\theta, \phi$) to the position ($\theta', \phi'$), the rotation operations are as follows: the rotation axis $\bm{n}$ is taken as $\bm{e}\times\bm{e}'$, and the rotation angle $\alpha$ is taken as $\arccos(\bm{e}\bullet\bm{e}')$. According to $\bm{n}$ and $\alpha$, the rotation matrix M$_r$ is constructed as:
\begin{equation}
\label{equation2}
M_r = \begin{bmatrix} n_x n_x (1-{\rm cos} \alpha) + {\rm cos} \alpha & n_x n_y (1-{\rm cos}\alpha) + n_z {\rm sin} \alpha & n_x n_z (1-{\rm cos} \alpha) - n_y {\rm sin} \alpha \\ 
n_x n_y (1-{\rm cos} \alpha) - n_z {\rm sin} \alpha & n_y n_y (1-{\rm cos} \alpha) + \rm cos \alpha & n_y n_z (1-{\rm cos} \alpha) + n_x {\rm sin} \alpha \\
n_x n_z (1-{\rm cos} \alpha) + n_y {\rm sin} \alpha & n_y n_z (1-{\rm cos} \alpha) - n_x \rm sin \alpha & n_z n_z (1-{\rm cos} \alpha) + {\rm cos} \alpha \\
\end{bmatrix}\\,
\end{equation}
where $\bm{n_i}$ is the projection of $\bm{n}$ on the $X$-, $Y$-, and $Z$-axes. Thus, it needs to be rotated once from ($\theta, \phi$) to ($\theta', \phi'$), and the $gimbal~lock$ problem can be avoided through the above operations. 
Then, we can adjust the rotation parameters ($\theta$, $\phi$) and ($\theta'$, $\phi'$) and the shock shape parameters ($h$, $s$, and $d$) until Equation (\ref{equation1}) best matches the shock wave fronts in the EUV observations of SDO and STEREO.
The wavelengths of 193 {\AA} and 211 {\AA} for AIA and 195 {\AA} for the EUVI are suitable for the observations of coronal shocks \citep{Ma2011,Su2015}. Therefore, the observations of these wavelengths are used to fit the shock surface.
The fitted surfaces of the shock at 09:24:42 and 09:26:18 UT are represented by the blue isolines in Figure \ref{fig:sh-fit2}. 

\subsection{Magnetic structure and $\theta_{\rm Bn}$ on the shock surface} \label{3.2}


Before determining the magnetic structure around the radio source, the uncertainty of the 3D location of the radio source is needed to be discussed. 
A conventional approach it to estimate the radial height of the radio source by the coronal density model \citep{Zucca2014b,Su2015,Morosan2019, Morosan2020}. 
The propagation effect of electromagnetic (EM) waves in the plasma is considered in this work. Since that the turbulent plasma can cause dispersion and scattering when EM waves propagate in plasmas\citep{Su2021}, the propagation effect may affect the apparent positions of the radio sources \citep{Kontar2017,Chen2020}. 
In this work, we use ray-tracing simulation results \citep{Zhang2021} to estimate the uncertainty of the radio source location. The refraction and anisotropic scattering effect of solar radio emission are considered in the calculation.
With the longitude of the radio source $\theta = 65^{\circ}$ (consistent with the eruption direction of the shock), the relative density fluctuation variance $\epsilon = 0.3$, and the anisotropic parameter $\alpha = 0.4$, we find that the location uncertainties of the radio source in the plane-of-the-sky at the harmonic frequencies 173.2 MHz and 150.9 MHz are about 0.024 and 0.023$\rm R_{\odot}$, respectively.

The coordinates of the centroids of the NRH radio sources at 173.2 and 150.9 MHz are (-1.12, -0.04)$\rm R_\odot$ and (-1.23, -0.17)$\rm R_\odot$ in the plane-of-the-sky, respectively. We take the beam size of the NRH observations as the location uncertainties. For the 173.2 MHz radio map, the location uncertainties are about $\pm0.09$$\rm R_{\odot}$ and $\pm0.22$$\rm R_{\odot}$ in the $x$ and $y$ axes in the plane-of-the-sky (see the left panel of Figure \ref{fig:nrh}), respectively;
For the 150.9 MHz radio map, the location uncertainties are about $\pm0.11$$\rm R_{\odot}$ and $\pm0.25$$\rm R_{\odot}$ in the $x$ and $y$ axes in the plane-of-the-sky (see the right panel of Figure \ref{fig:nrh}), respectively. 
In the Sun-Earth direction, we cannot get the radio centroids location from the NRH observations directly, considering the projection effect, we assume that the longitude of the radio source is approximate to that of the corresponding active region ($\theta=65^{\circ}$), thus, the radio centroids in the sun-earth direction are 0.47$\rm R_{\odot}$ and 0.52$\rm R_{\odot}$ at 173.2 and 150.9 MHz, respectively. We take the location uncertainty along the Sun-Earth direction to be the mean size of the beam in the plane-of-the-sky, which are about $\pm0.17$ and $\pm0.18$$\rm R_{\odot}$ at 173.2 and 150.9 MHz, respectively. In this way, the 3D uncertainties volume of the radio source location can be approximated as an elliptic cylinder. 
The radio source centriods with the location uncertainties in the plane-of-the-sky are shown in Figure \ref{fig:Mshock}, and the shock surface in the 3D uncertainty volume of the radio source at 173.2 and 150.9 MHz are shown as red shadows in Figures \ref{fig:thetaOnShock}b, c, e, f. We can see that most parts of the shock surface are in the 3D uncertainty volume of the radio source. 

The PFSS model is used to extrapolate the magnetic field around the shock front. Based on the extrapolated magnetic field, we can trace the magnetic field lines in the corona, and select the lines around the source region of the burst.
The magnetic field lines around the shock surface are shown as white and green lines in Figure \ref{fig:Mshock}, representing closed and open field, respectively.
Thus, the 3D magnetic structure around the shock front is constructed.
Since the PFSS model is a potential field model, it is only suitable for describing the magnetic field in the region undisturbed by the shock (upstream), but not the magnetic field in the region disturbed by the shock (downstream). 

Next, in order to distinguish whether the shock front at the source region of the radio burst is quasi-perpendicular or quasi-parallel, we need to describe the magnetic structure around the shock surface quantitatively. 
We overlay the magnetic field lines around the radio source region
onto the EUV images observed by SDO and STEREO. We can further obtain the intersection points between the fitted surface of the shock and the magnetic field lines. Then, we calculate the normal base vector (\bm{$e_{\rm n}$}) of the fitted shock surface and the tangent base vector (\bm{$e_{\rm t}$}) of the magnetic field lines at the intersection points. The angle $\theta_{\rm Bn}$ between \bm{$e_{\rm n}$} and \bm{$e_{\rm t}$} can be calculated as arccos($\bm{e_{\rm t}} \bullet \bm{e_{\rm n}}$). Usually, if the quantity of $\theta_{\rm Bn}$ is less than 45$^\circ$, the shock front at radio source is regarded as a quasi-parallel shock; if the quantity of $\theta_{\rm Bn}$ is larger than 45$^\circ$, the shock front is considered as a quasi-perpendicular shock. 
We sample the points on the shock surface evenly, and get $\theta_{\rm Bn}$ of these points on the shock surface. 

The distributions of $\theta_{\rm Bn}$ on the shock surface at 09:24:42 and 09:26:18 UT are shown in Figures \ref{fig:thetaOnShock}b and \ref{fig:thetaOnShock}e, and the nose of the shock is indicated by the white asterisks. 
We also plot the histograms of $\theta_{\rm Bn}$ of the radio source at 09:24:42 and 09:26:18 UT in Figure \ref{fig:hist}. We find that
the distributions of $\theta_{\rm Bn}$ can be roughly separated into two parts, one for $\theta_{\rm Bn}\lesssim45^{\circ}$, the other for $\theta_{\rm Bn}\gtrsim45^{\circ}$, meaning that $\theta_{\rm Bn}$ has an obvious bimodal distribution of $\theta_{\rm Bn}$. 
The bimodal distribution of $\theta_{\rm Bn}$ implies that both quasi-parallel and quasi-perpendicular shocks may be at work for the generation of the type II radio burst.
The average values of $\theta_{\rm Bn}$ of the radio sources on the shock surface are 36$\pm$16$^{\circ}$ and 42$\pm$20$^{\circ}$ at 09:24:42 and 09:26:18 UT, respectively.

The magnetic field strength at the shock upstream ($B_1$) is obtained by the PFSS model here. The distributions of $B_1$ on the shock surface at 09:24:42 and 09:26:18 UT are shown in Figures \ref{fig:thetaOnShock}c and \ref{fig:thetaOnShock}f, respectively. The mean values of $B_1$ of the radio source on the shock surface are 3.05$\pm$0.74 and 1.70$\pm$0.39 Gauss at 09:24:42 and 09:26:18 UT, respectively. 
The histograms of $B_1$ of the radio source at 09:24:42 and 09:26:18 UT are shown in the second row of Figure \ref{fig:hist}.
It is seen that, 
different from the bimodal distributions of $\theta_{\rm Bn}$, the value of $B_1$ shows roughly a unimodal distribution but with asymmetry. 

\subsection{Alfv\'{e}n Mach number of the shock}

The Alfv\'{e}n Mach number, $M_A = v/v_A$, is defined as the ratio between the speed $v$ of the shock and the Alfv\'{e}n speed $v_A$.
$v_A$ is expressed as
\begin{equation}
\label{equation3}
v_A = \frac{B_1}{\sqrt{\mu_0 \rho}}, \\
\end{equation}
where $\mu_0$ is the magnetic permeability of the vacuum, $B_1$ and $\rho$ are the magnetic field strength and the mass density of the shock upstream \citep{Priest2014}. 
Here, we expect to get the distributions of $v_A$ and $M_A$ on the shock front. 
As mentioned in Section 3.2, the distribution of $B_1$ on the shock surface can be derived from the PFSS model, and the results are shown in Figure \ref{fig:thetaOnShock}c and \ref{fig:thetaOnShock}f.

Besides the distribution of $B_1$, we also need to obtain the distribution of $\rho$ in order to estimate the distribution of $v_A$ on the shock surface.
In theory, the fundamental frequency of the type II radio bursts can be deemed as the Langmuir frequency of the shock upstream \citep{Cairns1985,Cairns1986,Cairns1988,Knock2005}. With the relation between the Langmuir frequency $f$ and the number density $n_{\rm e}$, $f = 8980 \sqrt{n_{\rm e}}$, we can then derive the value of $n_{\rm e}$. 
Note that the frequencies corresponding to the red markers in Figure \ref{fig:spec} are the first harmonic frequency of the type II radio burst, thus the corresponding fundamental frequencies are approximately 173.2/2 and 150.9/2 MHz, and the corresponding $n_{\rm e}$ is $9.3\times10^7$ and $7.0\times10^7~\rm cm^{-3}$, respectively. 

As most of the coronal density models are only in the radial direction, \citep[e.g.,][]{Newkirk1961,Mann1999}, the density variations in the longitude and latitude directions are not available. 
\citet{Zucca2014a} used the DEM method to calculate 2D distributions of $n_{\rm e}$ around the shock background in the corona. \citet{Rouillard2016} further updated this approach to esimate the 3D the distributions of $n_{\rm e}$ around the shock background. This powerful approach has also been applied in other works \citep{Zucca2018,Frassati2019}. 
Based on the observations by SDO/AIA at 6 EUV wavelengths, i.e., 93, 131, 171, 193, 211, 335 \AA, we can use the DEM method to derive the distributions of $n_{\rm e}$ in the corona \citep{Aschwanden2001}. A widely used DEM method proposed by \cite{Weber2004} is applied \citep{Cheng2012,Su2016,Su2018} in this work. 
According to the approach of \citet{Rouillard2016}, we obtain the distributions of $n_{\rm e}$ in the corona every 6 hours by the DEM method, which is equivalent to rotating the corona in the plane-of-the-sky every 3.3$^{\circ}$ each time. From March 3, 2014 to March 5, 2014, we repeat the process mentioned above 8 times, covering the entire volume of the shock surface shown in Figure \ref{fig:thetaOnShock}, which spans nearly 30$^{\circ}$ in longitude. 
Then, based on the distributions of $n_{\rm e}$ on the 8 meridional planes, we get the distribution of $n_{\rm e}$ on the shock surface by interpolation, and the results are shown in Figures \ref{fig:MA}a and \ref{fig:MA}d.
The mean values of $n_e$ of the radio source on the shock surface are 10.0$\pm0.6\times10^7$ and 7.9$\pm0.2\times10^7$ cm$^{-3}$ at 09:24:42 and 09:26:18 UT, respectively. 

Based on the distributions of $B_1$ and $\rho$, we can estimate the distribution of $v_A$ on the shock surface by Equation (\ref{equation3}). The results are shown in Figures \ref{fig:MA}b and \ref{fig:MA}e. 
Similar to the distributions of $B_1$, the distribution of $v_A$ is smooth on the shock surface. The mean values of $v_A$ of the radio source on the shock surface are 585$\pm$156 and 366$\pm$88 km s$^{-1}$ at 09:24:42 and 09:26:18 UT, respectively. 
The histograms of $v_A$ on the shock surface are shown in the fourth row of Figure \ref{fig:hist}, 
which roughly shows a skewed unimodal distribution.

\begin{figure}[ht]
	\centering
	\begin{minipage}[b]{\textwidth}
		\centering
		\includegraphics[width=12cm]{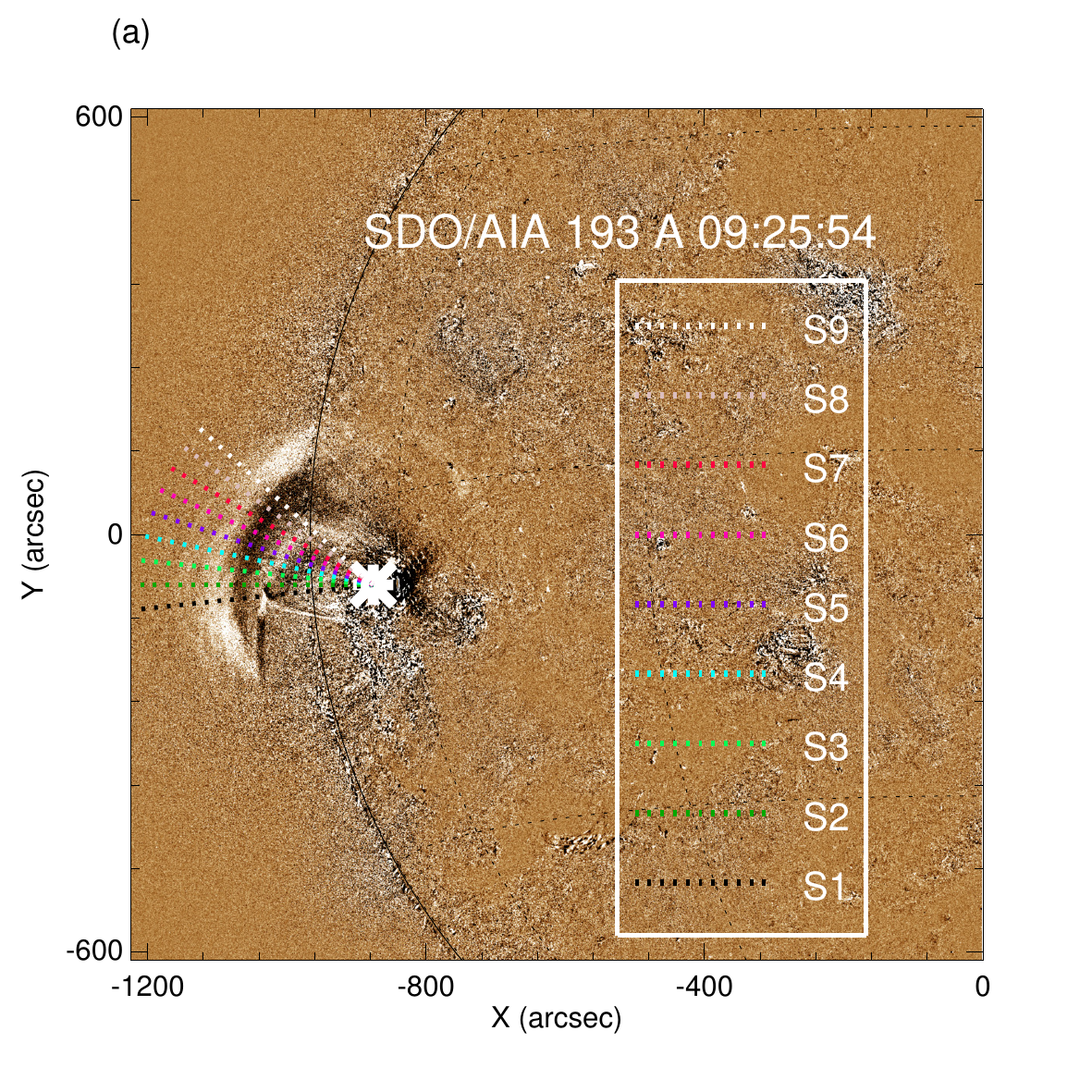}
	\end{minipage}
	\begin{minipage}[b]{\textwidth}
		\centering
		\includegraphics[width=12cm]{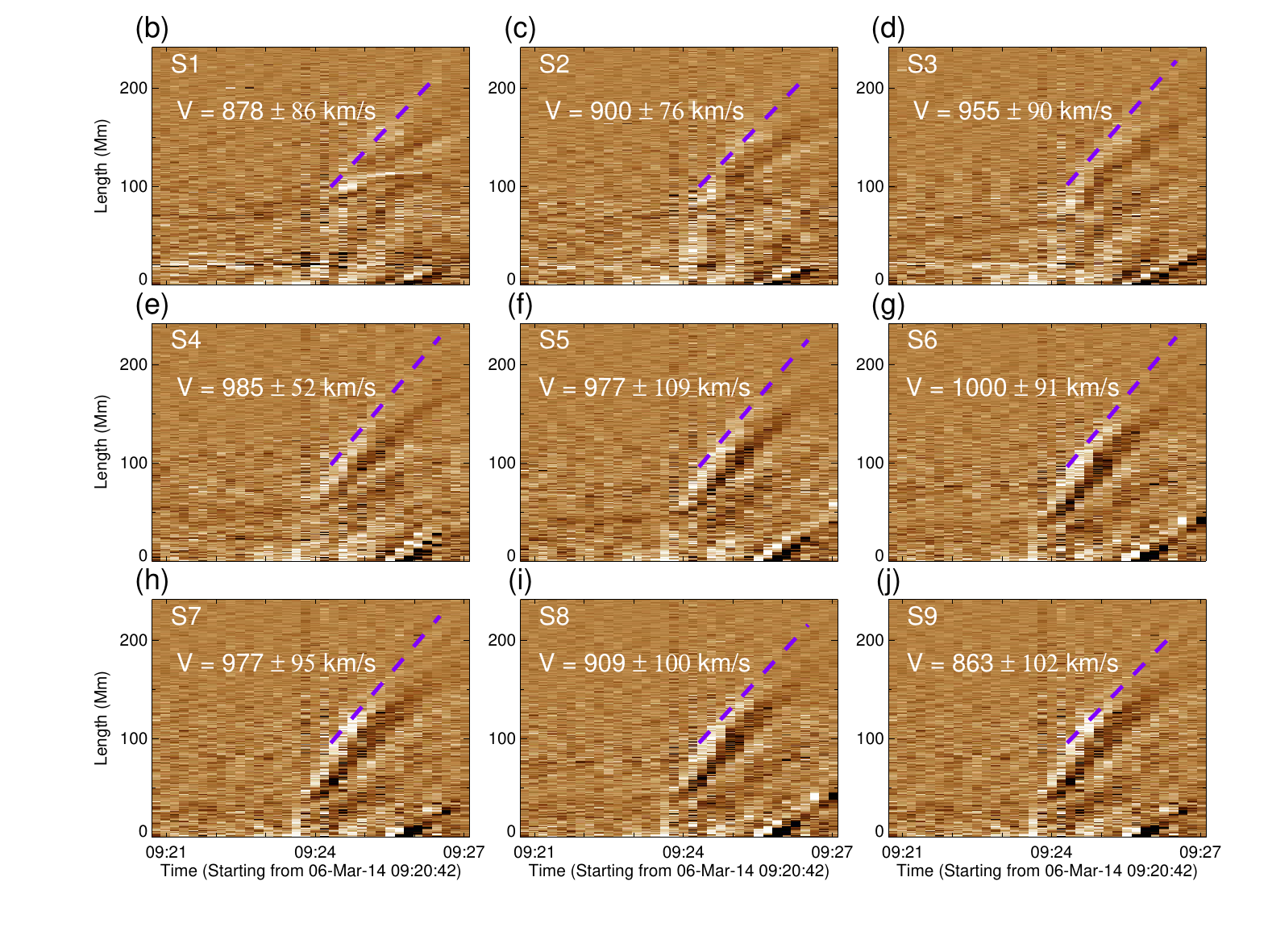}
	\end{minipage}
	\caption{Panel (a) is the running difference image of SDO/AIA 193 \AA~ between 09:25:54 and 09:25:30 UT. 
		There are 9 colored slices along the different directions around the shock surface.
		In the clockwise direction, these 9 slices are denoted as S1, S2, S3, S4, S5, S6, S7, S8 and S9.
	    Panels (b, c, d, e, f, g, h, i, j) are the time-distance diagrams of the slices S1--S9 in panel (a), respectively.
			}
	\label{fig:EUV-TD}
\end{figure}

\begin{figure}[ht]
	\begin{minipage}[b]{\textwidth}
		\centering
		\includegraphics[width=12cm]{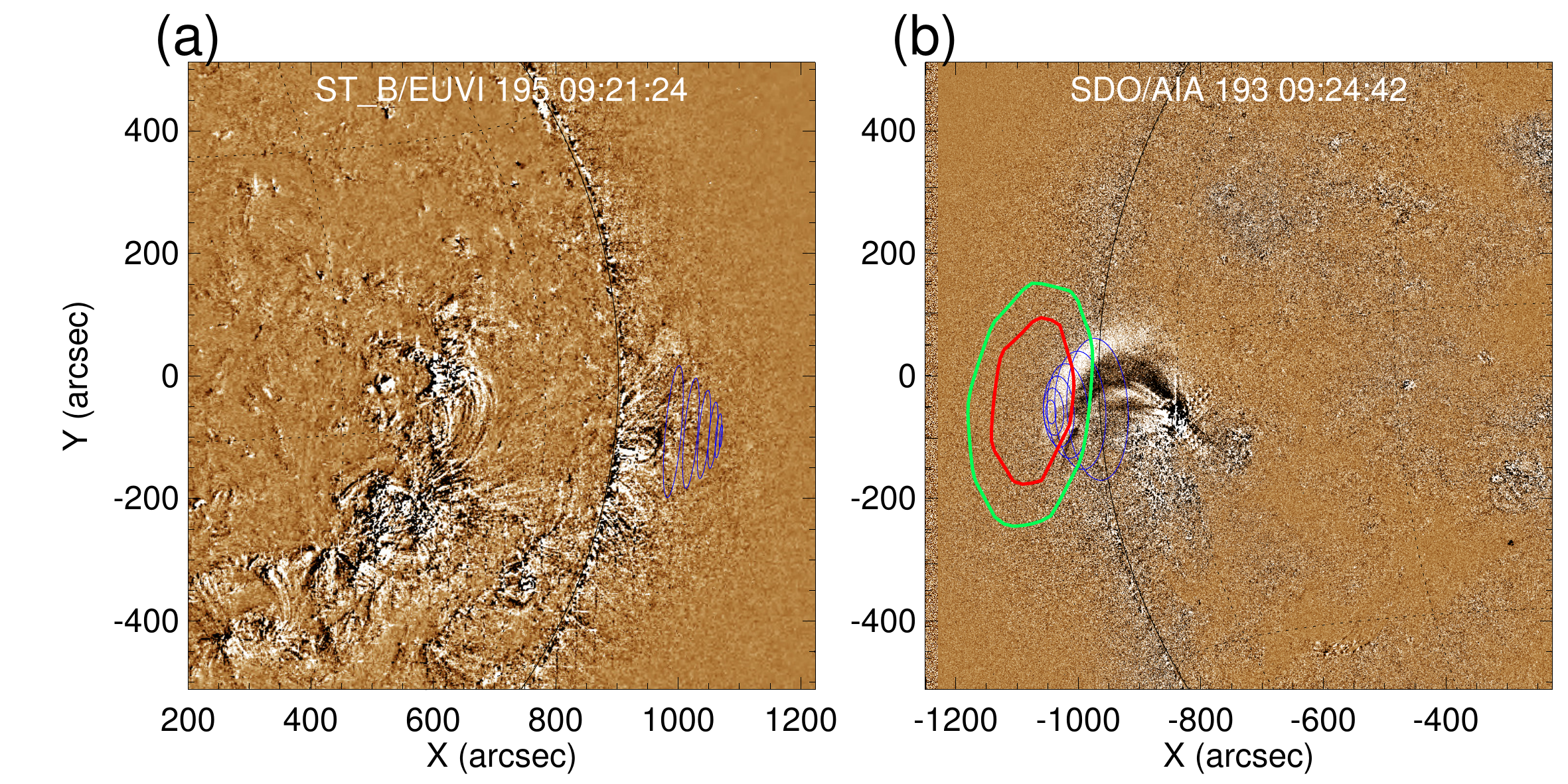}
	\end{minipage}
	\begin{minipage}[b]{\textwidth}
		\centering
		\includegraphics[width=12cm]{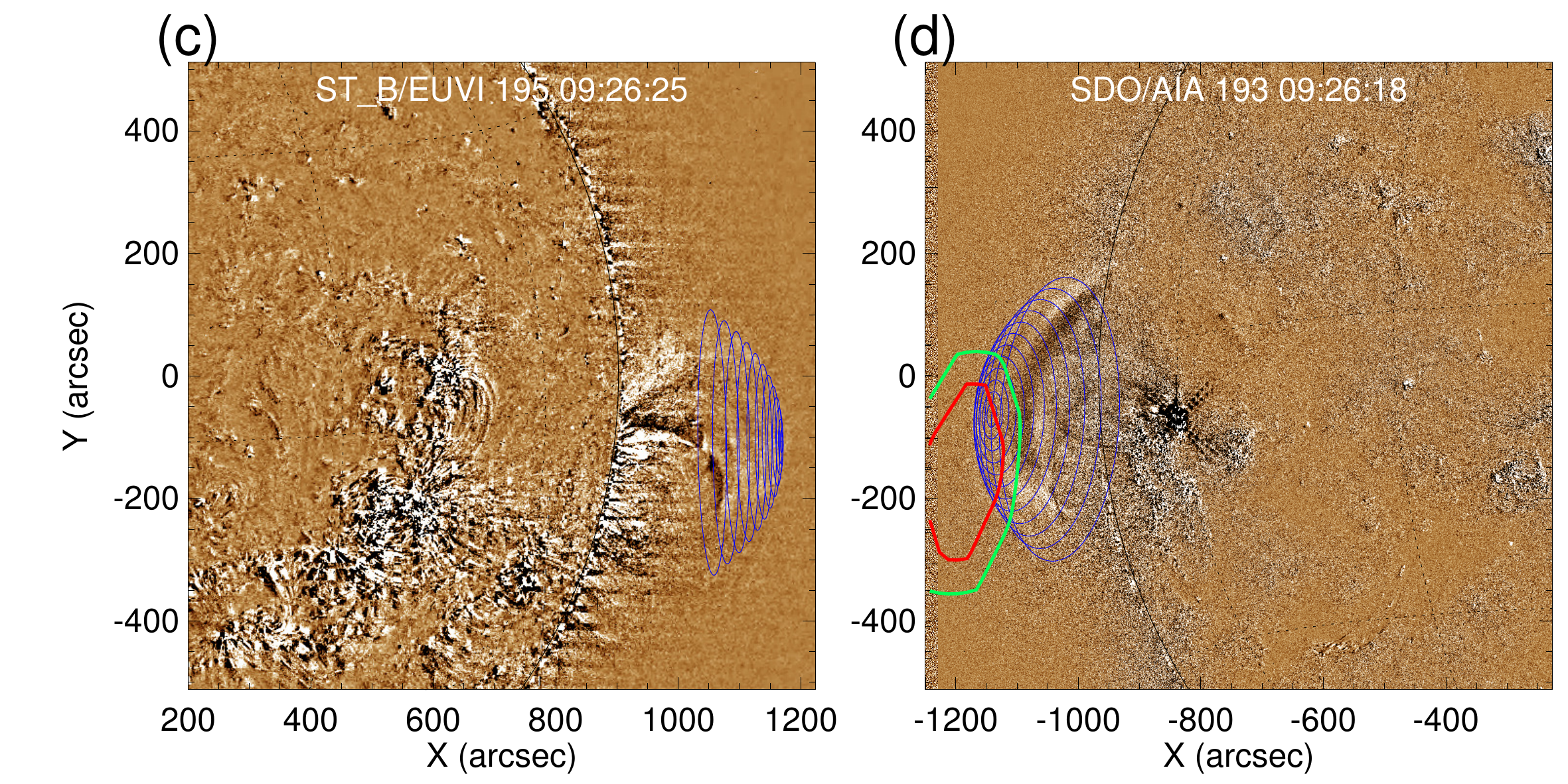}
	\end{minipage}
	\caption{Evolution of the coronal shock at 193 {\AA} of SDO/AIA and 195 {\AA} of ST\_B/EUVI. 
		The blue isolines overlaied on SDO/AIA and ST\_B/EUVI illustrate the 3D fitted surface of the shock. The radio sources at 173.2 (09:24:42 UT) and 150.9 MHz (09:26:18 UT) are shown in panel (b) and (d), respectively. Green and red contours represent brightness temperature levels at 80\% and 90\% of their respective maximums.}
	\label{fig:sh-fit2}
\end{figure}

\begin{figure}[ht]
	\centering
	\includegraphics[width=12cm]{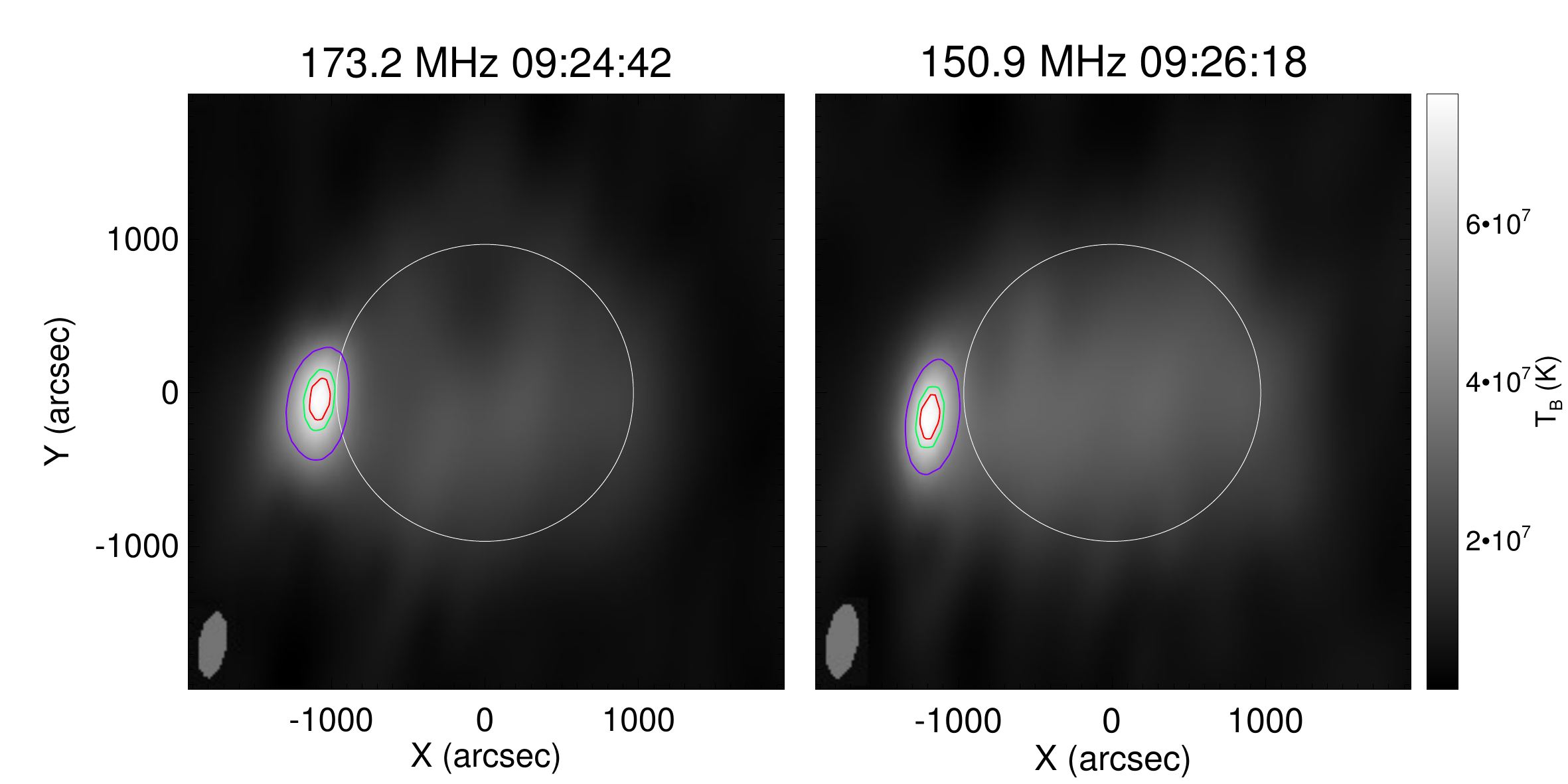}
	\caption{The radio source of the type II radio burst at 173.2 (09:24:42 UT) and 150.9 MHz (09:26:18 UT). The purple, green and red contours are 50\%, 80\% and 90\% of the brightness temperature maximum.}
	\label{fig:nrh}
\end{figure}

\begin{figure}[ht]
	\begin{minipage}[b]{\textwidth}
		\centering
		\includegraphics[width=12cm]{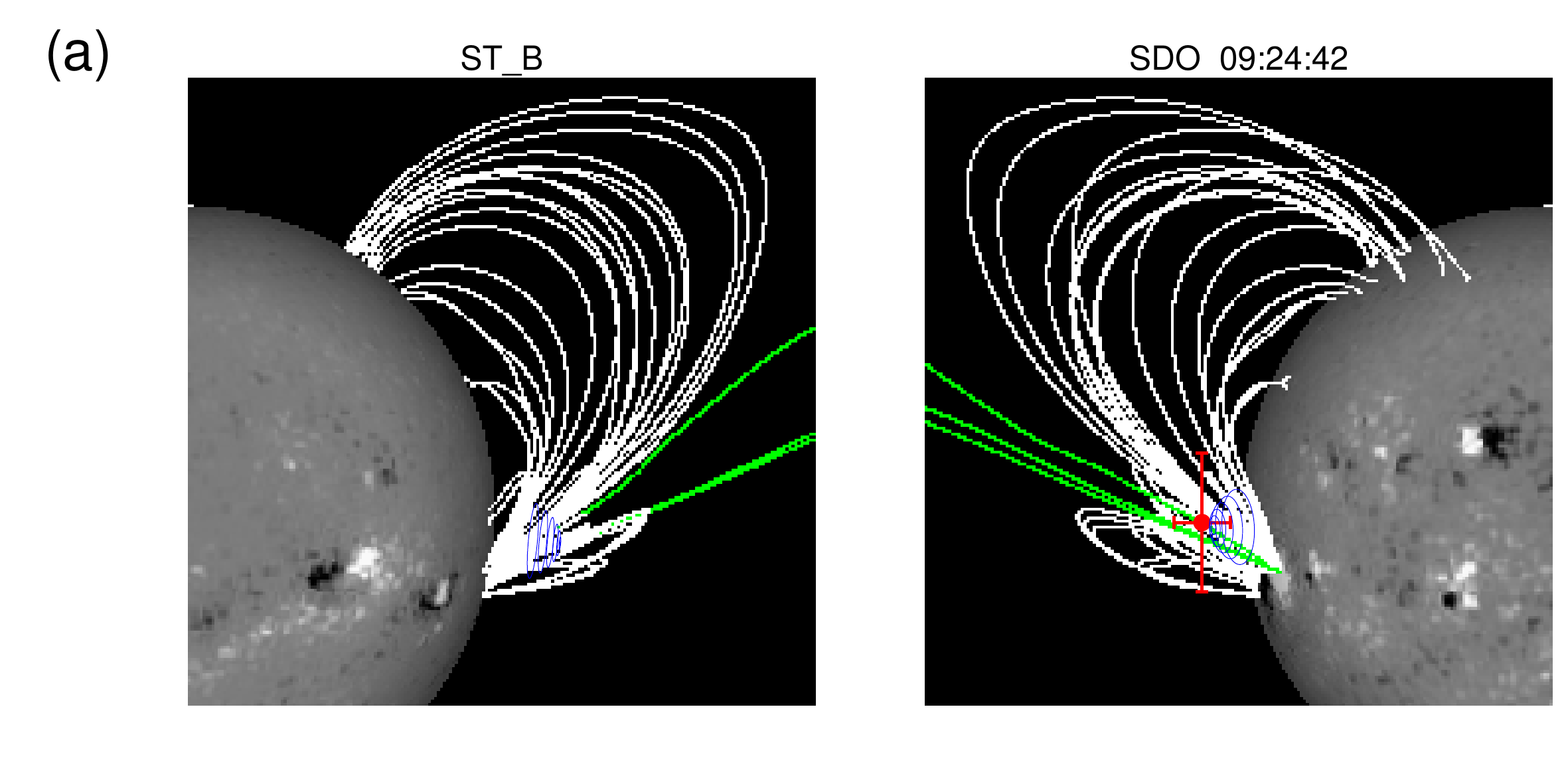}
	\end{minipage}
	\begin{minipage}[b]{\textwidth}
		\centering
		\includegraphics[width=12cm]{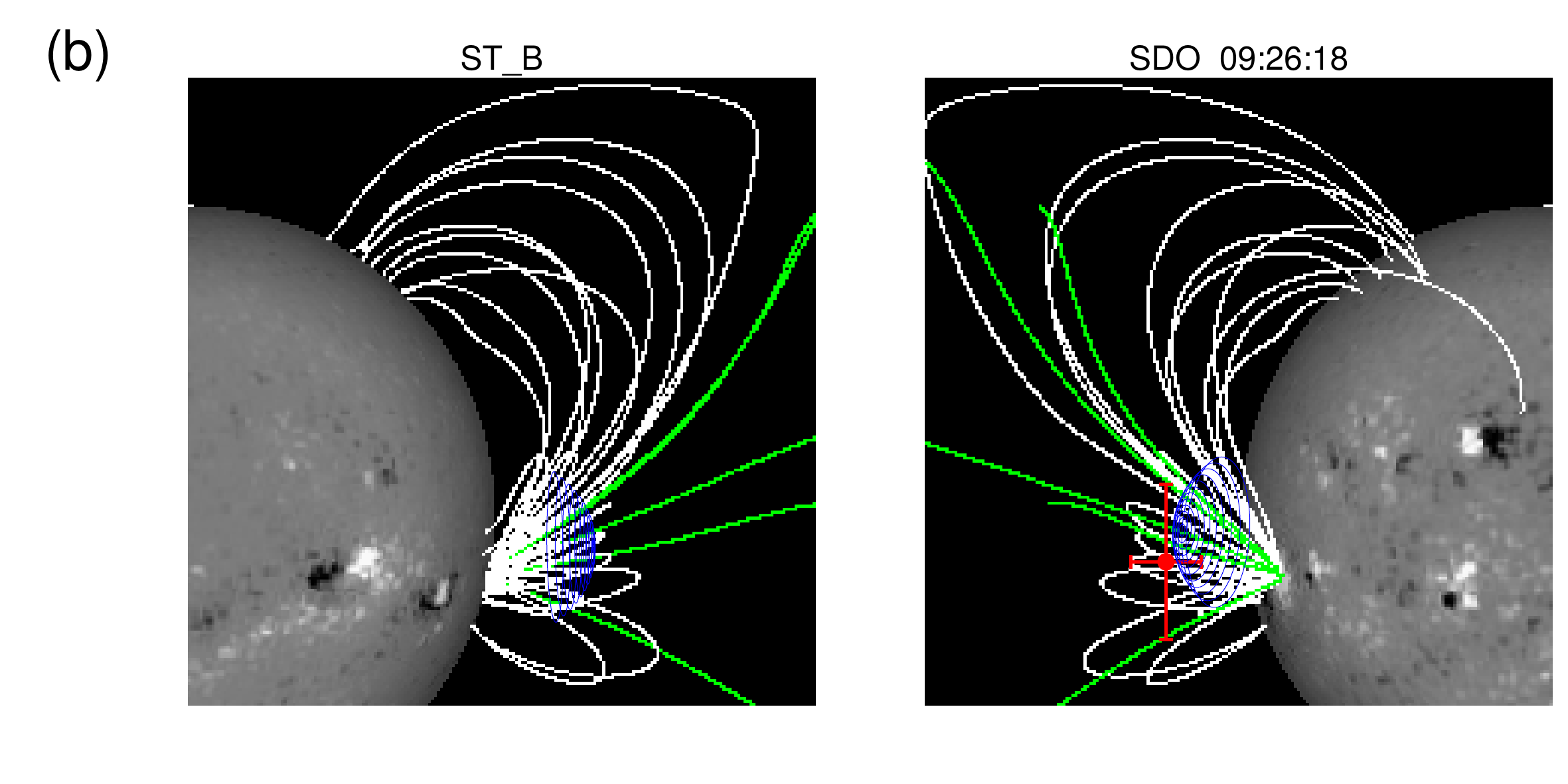}
	\end{minipage}
	\caption{The magnetic structure around the shock surface. 
		The white and green lines represent the closed and open field lines which extrapolated by the PFSS model, respectively. 
		The blue isolines in panels (a) and (b) are the fitted surface of the shock at 09:24:42 and 09:26:18 UT, respectively, which are the same as that in Figure \ref{fig:sh-fit2}.
		The red solid circles represent the radio centroids at 09:24:42 and 09:26:18 UT, respectively.
		}
	\label{fig:Mshock}
\end{figure}

\begin{figure}[ht]
	\begin{minipage}[b]{\textwidth}
		\centering
		\begin{minipage}[b]{0.2\textwidth}
			\includegraphics[width = 4.5 cm,trim = 0 0 0 0,clip = true]{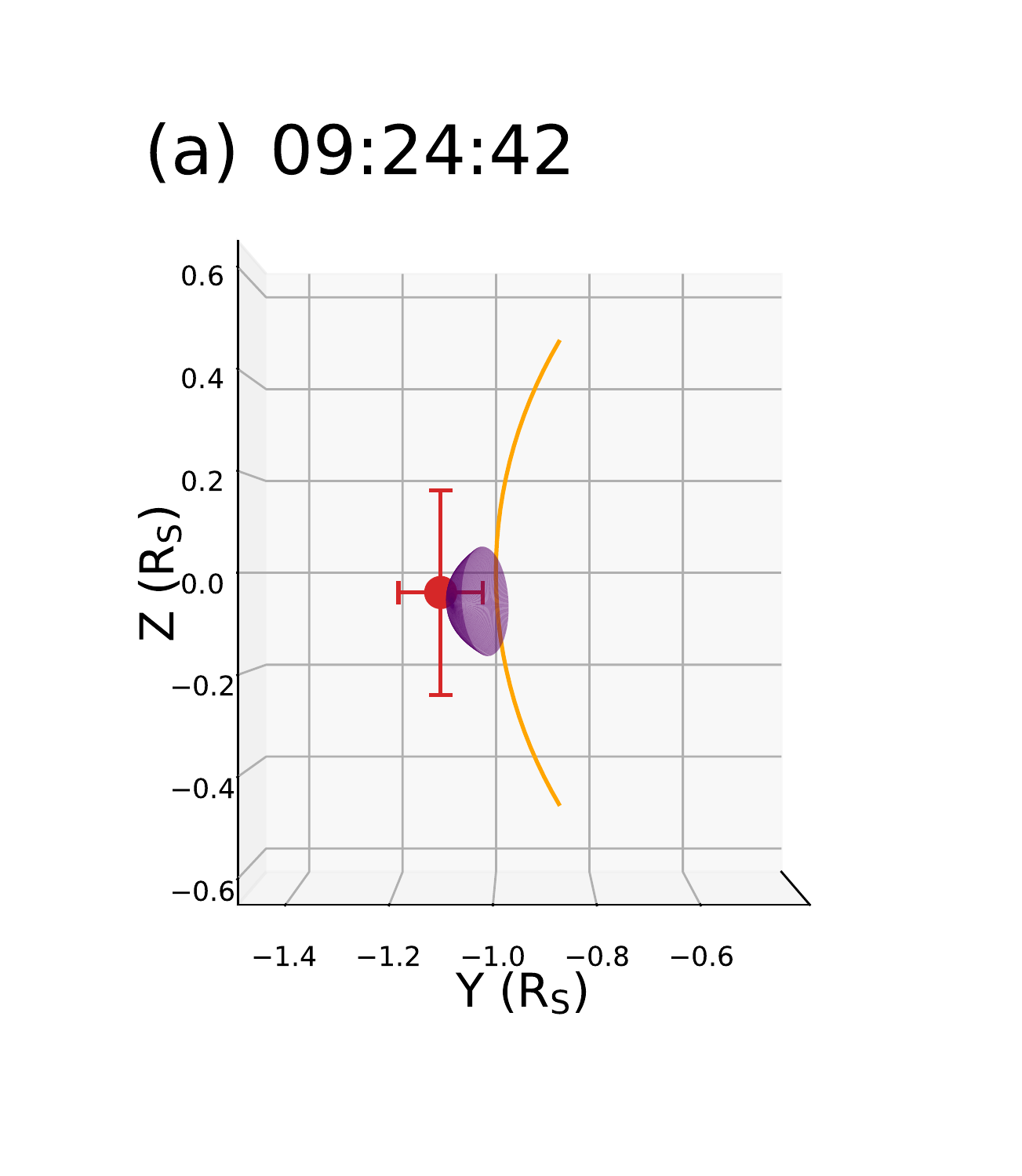}
		\end{minipage}
		\begin{minipage}[b]{0.32\textwidth}
			\includegraphics[width = 6.25 cm,trim = 0 0 0 0,clip = true]{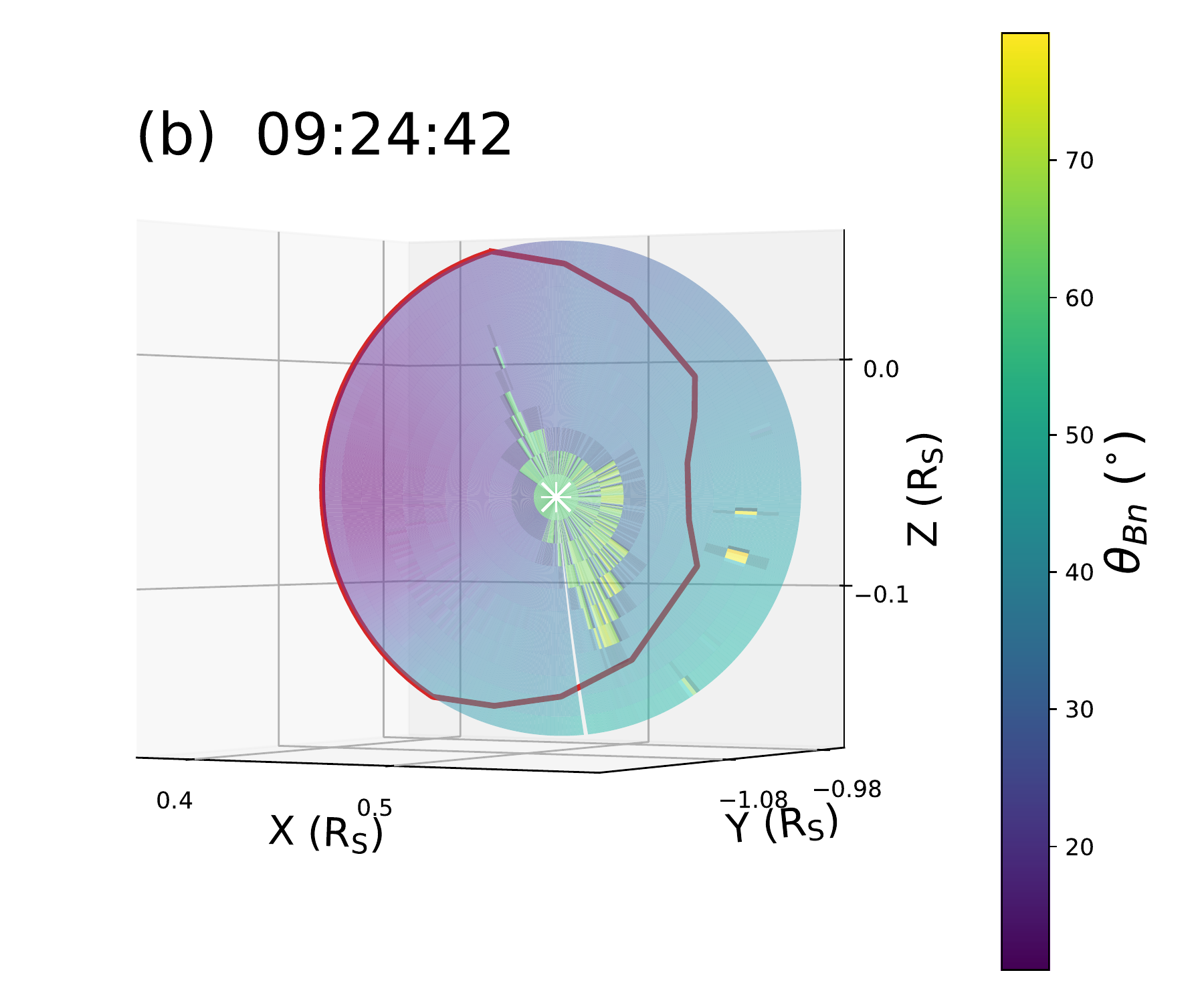}
		\end{minipage}
		\begin{minipage}[b]{0.32\textwidth}
			\includegraphics[width = 6.25 cm,trim = 0 0 0 0,clip = true]{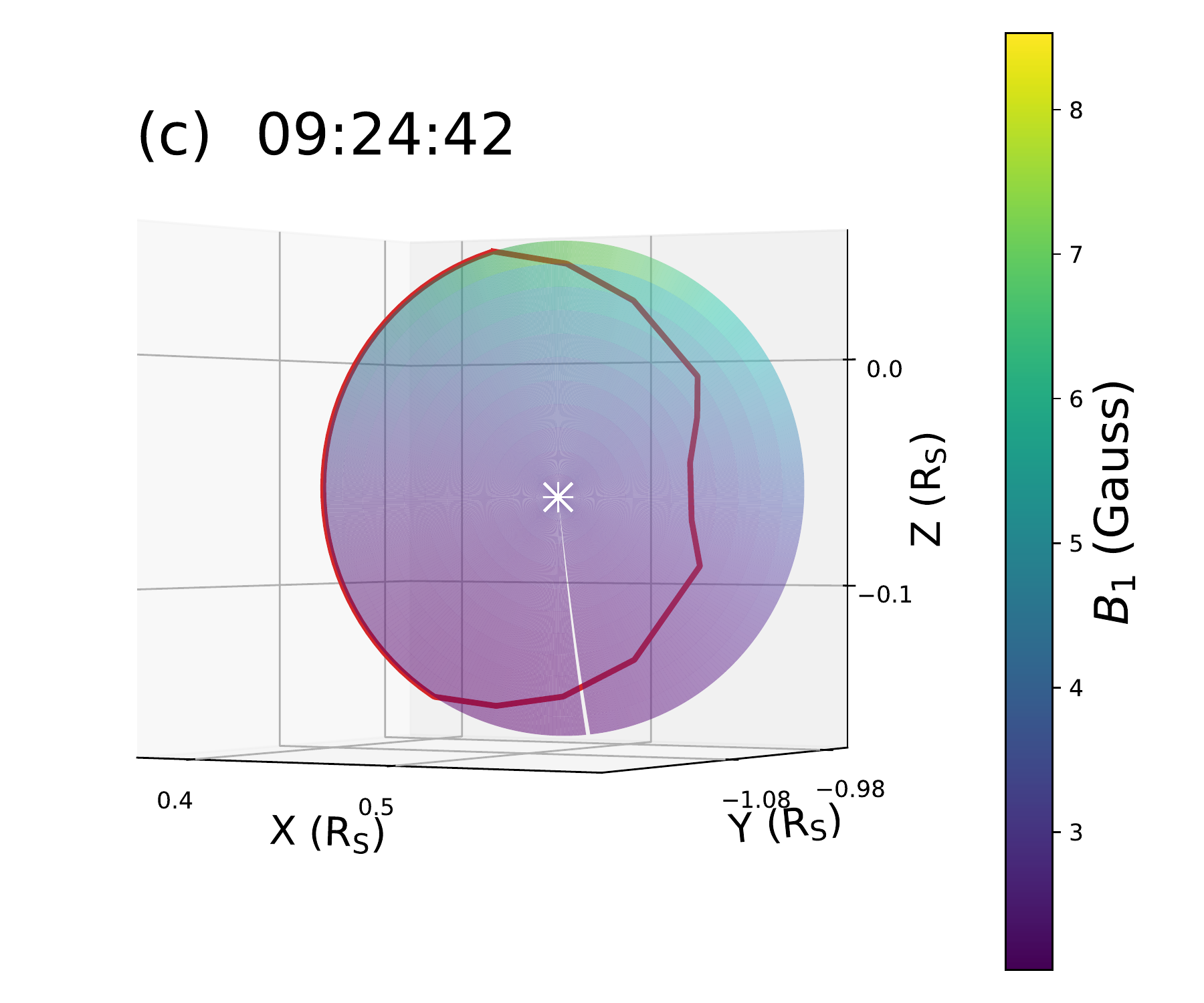}
		\end{minipage}
	\end{minipage}
	\begin{minipage}[b]{\textwidth}
		\centering
		\begin{minipage}[b]{0.2\textwidth}
			\includegraphics[width = 4.5 cm,trim = 0 0 0 0,clip = true]{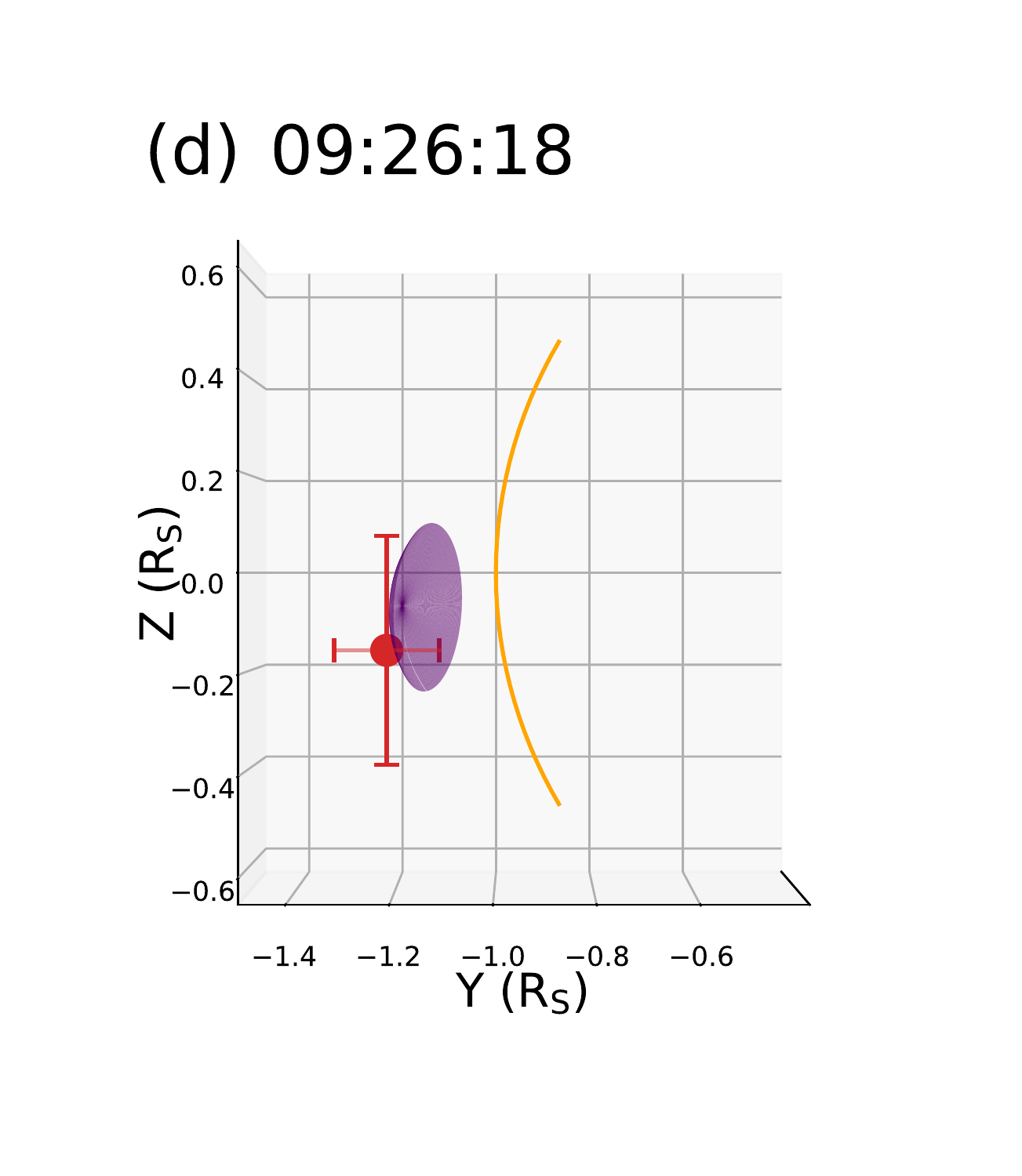}
		\end{minipage}
		\begin{minipage}[b]{0.32\textwidth}
			\includegraphics[width = 6.25 cm,trim = 0 0 0 0,clip = true]{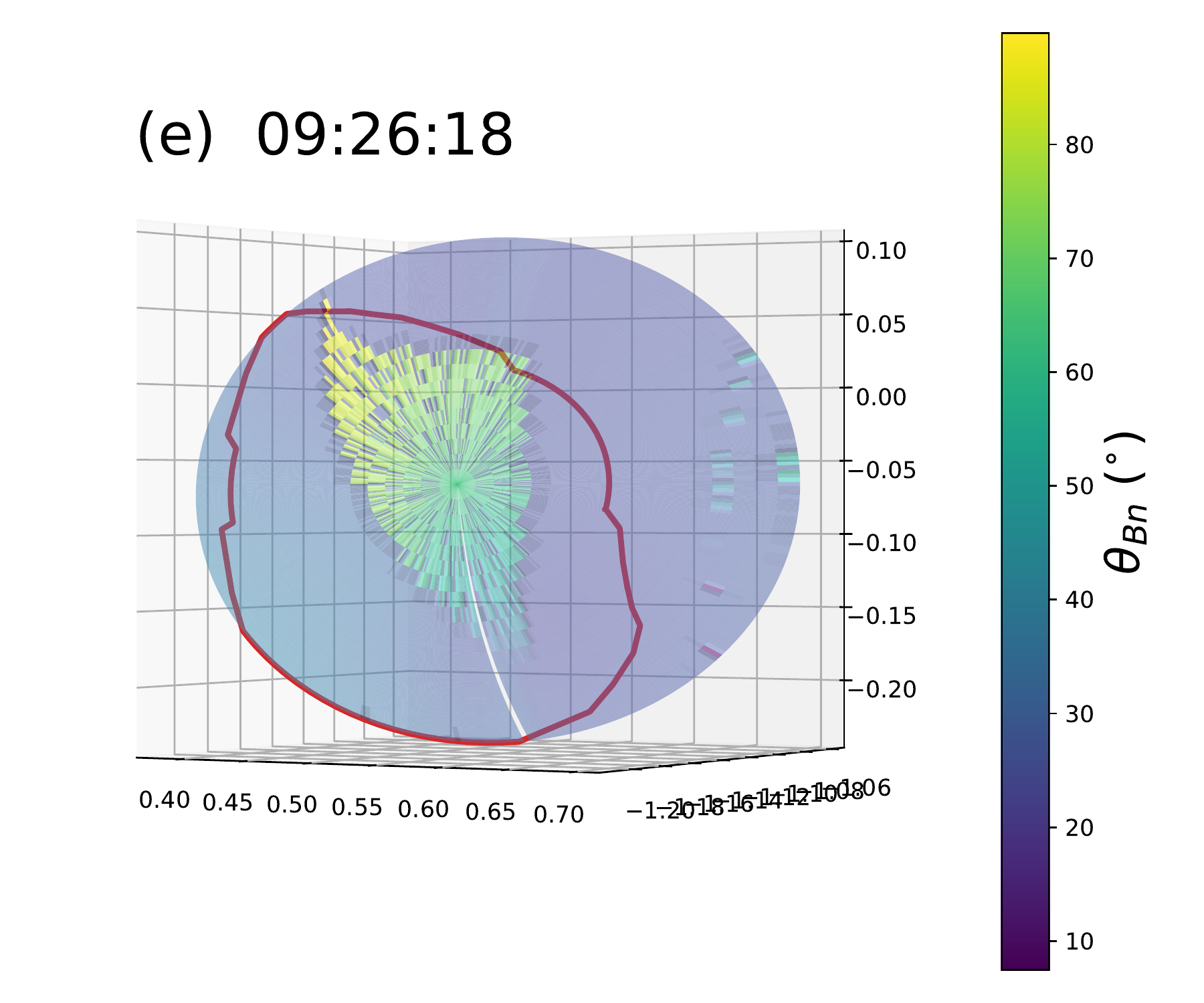}
		\end{minipage}
		\begin{minipage}[b]{0.32\textwidth}
			\includegraphics[width = 6.25 cm,trim = 0 0 0 0,clip = true]{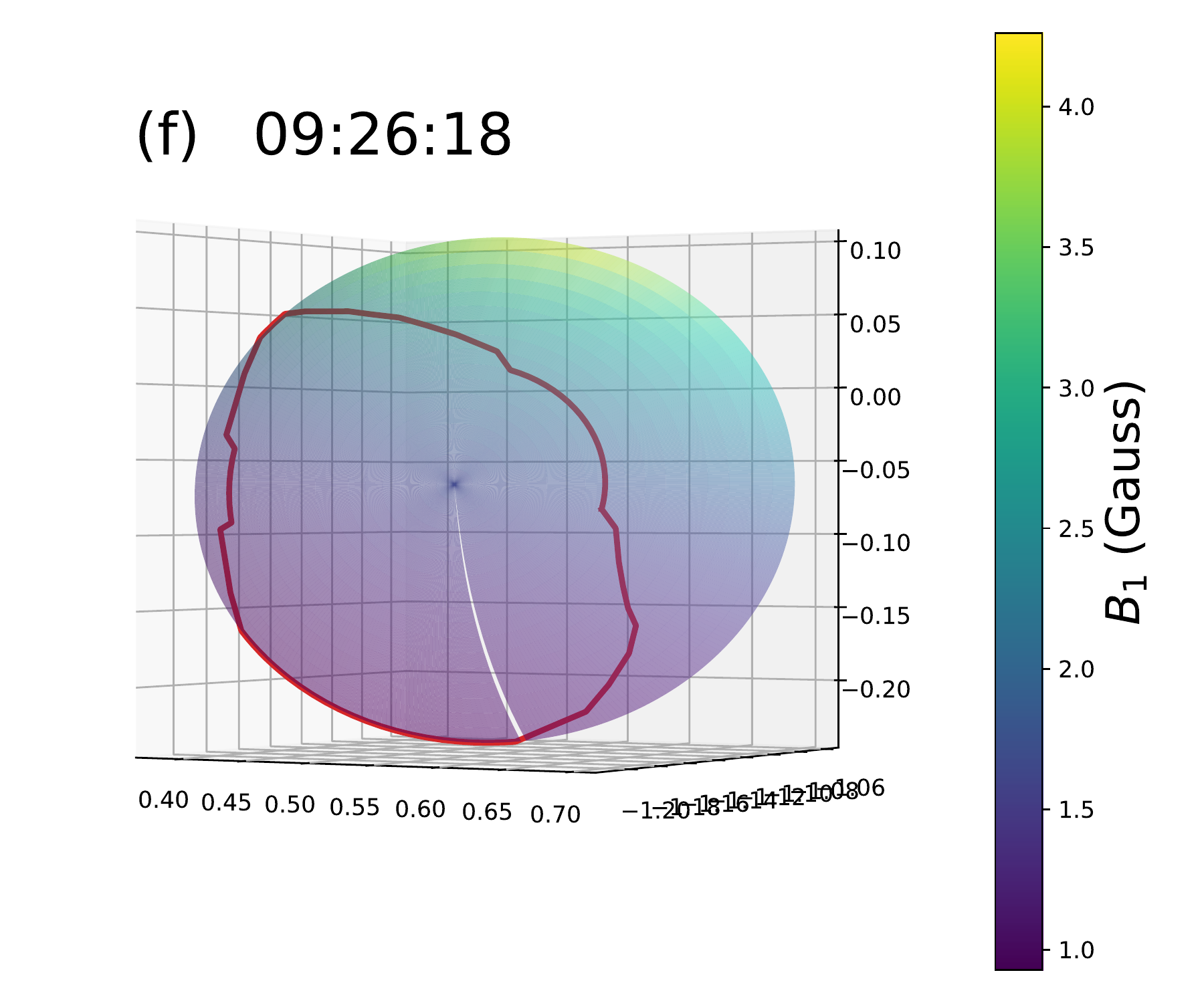}
		\end{minipage}
	\end{minipage}
	\caption{The distribution of $\theta_{\rm Bn}$ and the value of upstream magnetic field ($B_1$) on the shock surface.
		The panels (a) and (d) show the shock surfaces with a viewing perspective from SDO in the heliocentric coordinate system at 09:24:42 and 09:26:18 UT, respectively, the red solid circles represent the radio centroids, the solar limb is represented by the orange arc. 
				The upper (b, c) and lower (e, f) panels are the distribution of $\theta_{\rm Bn}$ and $B_1$ at 09:24:42 and 09:26:18 UT, respectively. 
				The panels (b) and (e) show the details of $\theta_{\rm Bn}$ on the shock surfaces, the quantities of $\theta_{\rm Bn}$ are indicated by the colorbar. 
				The panels (c) and (f) show the distributions of $B_1$ on the shock surfaces, the quantities of $B_1$ are indicated by the colorbar. 
				The radio sources on the shock surface are shown as red contours in panels (b), (c), (e), and (f).
				The white asterisks are the noses of the shock surface at 09:24:42 and 09:26:18 UT.}
	\label{fig:thetaOnShock}
\end{figure}

In order to derive the distribution of $M_A$ on the shock surface, we need to get the shock speed $v_{sh}$ first.
We select 9 slices starting from the eruption source site (marked as a white asterisk in Figure 3) across the shock front, and the neighboring slices are separated by $6^{\circ}$, as shown in Figure \ref{fig:EUV-TD}a.
In the clockwise sequence, we denote these 9 slices as S1, S2, S3, S4, S5, S6, S7, S8 and S9. 
The time-distance diagrams along the slices S1--S9 are displayed in the bottom panels of Figure \ref{fig:EUV-TD},
from which we measure the shock speed $v$.
The shock speeds along the 9 directions are 878$\pm$86 km s$^{-1}$, 900$\pm$76 km s$^{-1}$, 955$\pm$76 km s$^{-1}$, 985$\pm$52 km s$^{-1}$, 977$\pm$109 km s$^{-1}$, 1000$\pm$91 km s$^{-1}$, 977$\pm$95 km s$^{-1}$, 909$\pm$100 km s$^{-1}$, and 863$\pm$102 km s$^{-1}$, respectively. 

The speeds we measured from the EUV images are the speeds of the outermost edge of the 3D shock surface along 9 slices projected in the plane-of-the-sky. 
Obviously, in 3D space, the outermost edge is where the line-of-sight is tangential to the shock surface.
For each slice, we find the heliocentric coordinates ($x, y, z$) of each tangent point of the line-of-sight and the 3D shock surface (tangent point). 
Taking into account the projection effect, and combining the shock speed ($v_{sh}$) measured from the time-distance diagrams (bottom panels of Figure \ref{fig:EUV-TD}), we get the revised shock speeds, $v$, along each direction based on the heliocentric coordinates ($x, y, z$) of each points.
According to the shape of the shock surface (Equation (\ref{equation1})) and the coordinate transformation relationship (Equation (\ref{equation2})), we can get the value of $h$ (defined in Equation (\ref{equation1})) for each intersection points through its heliocentric coordinates ($x, y, z$) of each tangent points. We assume that the shock speed of the points on the shock surface with the same $h$ are the same. Then through linear interpolation, we can get the distribution of shock speed on the shock surface.

With the distributions of $v$ and $v_A$, we can estimate the distribution of $M_A = v/v_A$ on the shock surface, and the results are shown in Figures \ref{fig:MA}c and \ref{fig:MA}f.
Similar to the distributions of $B_1$, and $v_A$, the distribution of $M_A$ is smooth on the shock surface as well. The mean values of $M_A$ of the radio source on the shock surface are 1.64$\pm$0.32 and 2.61$\pm$0.47 at 09:24:42 and 09:26:18 UT, respectively. 
The histograms of $M_A$ of the radio source on the shock surface at 09:24:42 and 09:26:18 UT are shown in Figure \ref{fig:hist}.
Similarly, the distributions of $M_A$ show a roughly unimodal distribution with asymmetry.

\begin{table}[]
	\caption{Mean value and standard deviation (SD) coefficients of $\theta_{\rm Bn}$, $B_1$, $n_e$, $v_A$, $M_A$, $X$, and $B_2$ of the radio source at 09:24:42 UT and 09:26:18 UT.
	}
	\centering
	\begin{tabular}{c|cc|cc}
		\hline
		& \multicolumn{2}{c|}{09:24:42 UT}                        & \multicolumn{2}{c}{09:26:18 UT}                         \\ \hline
		& mean                            & SD coefficient & mean                            & SD coefficient \\ \hline
		$\theta_{\rm Bn}~(^{\circ})$ & 36 $\pm$ 16                      & 0.42             & 42 $\pm$ 20                           & 0.47             \\ \hline
		$B_1$ (Gauss)                  & 3.05 $\pm$ 0.74                       & 0.35            & 1.70 $\pm$ 0.39                       & 0.33              \\ \hline
		$n_e$ ($\rm cm^{-3}$)        & $10.0 \pm 0.6 \times 10^7$ & 0.06            & $7.9 \pm 0.2 \times 10^7$ & 0.03            \\ \hline
		$v_A$ (km/s)                 & 585 $\pm$ 156                          & 0.27            & 366 $\pm$ 88                          & 0.24             \\ \hline
		$M_A$                        & 1.64 $\pm$ 0.32                        & 0.19            & 2.61 $\pm$ 0.47                        & 0.18              \\ \hline
		$X$                        & 2.10 $\pm$ 0.53                        & 0.25            & 2.99 $\pm$ 0.45                        & 0.15              \\ \hline
		$B_2$ (Gauss)                        & 6.25 $\pm$ 0.89                        & 0.14            & 4.09 $\pm$ 0.98                        & 0.24              \\ \hline
	\end{tabular}
	\label{table1}
\end{table}

\subsection{Properties of shock downstream}

\begin{figure}[ht]
	\centering
		\begin{minipage}[b]{0.411\textwidth}
			\centering
			\includegraphics[width = 6 cm]{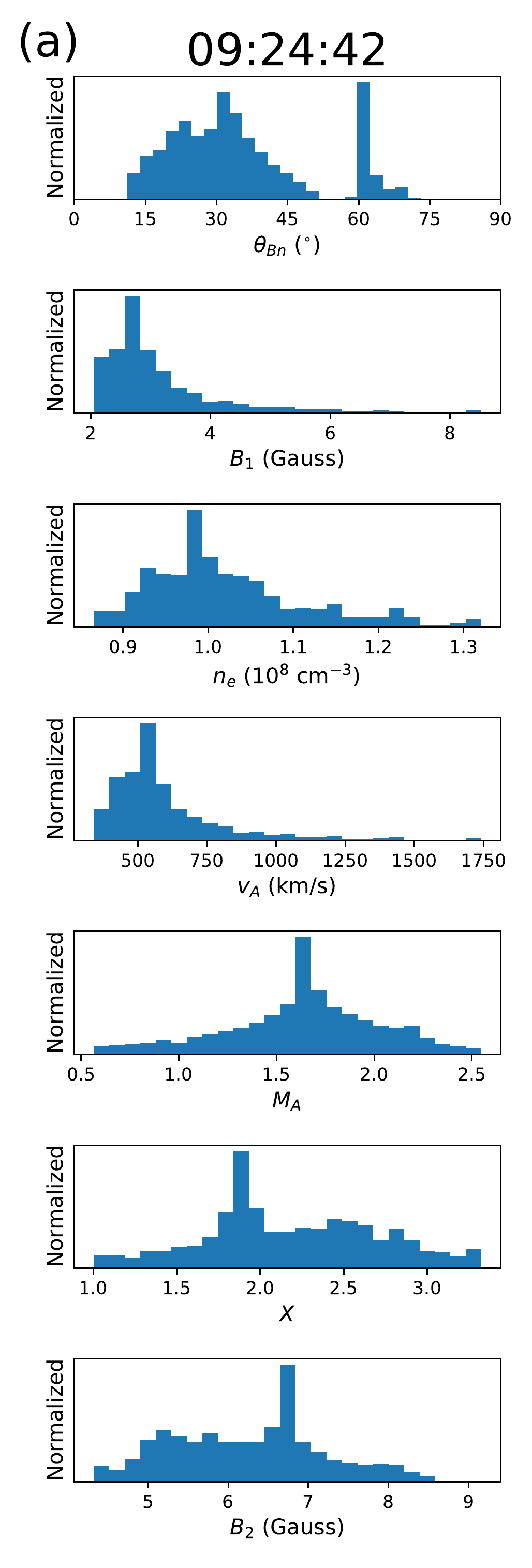}
		\end{minipage}
		\begin{minipage}[b]{0.411\textwidth}
			\centering
			\includegraphics[width = 6 cm]{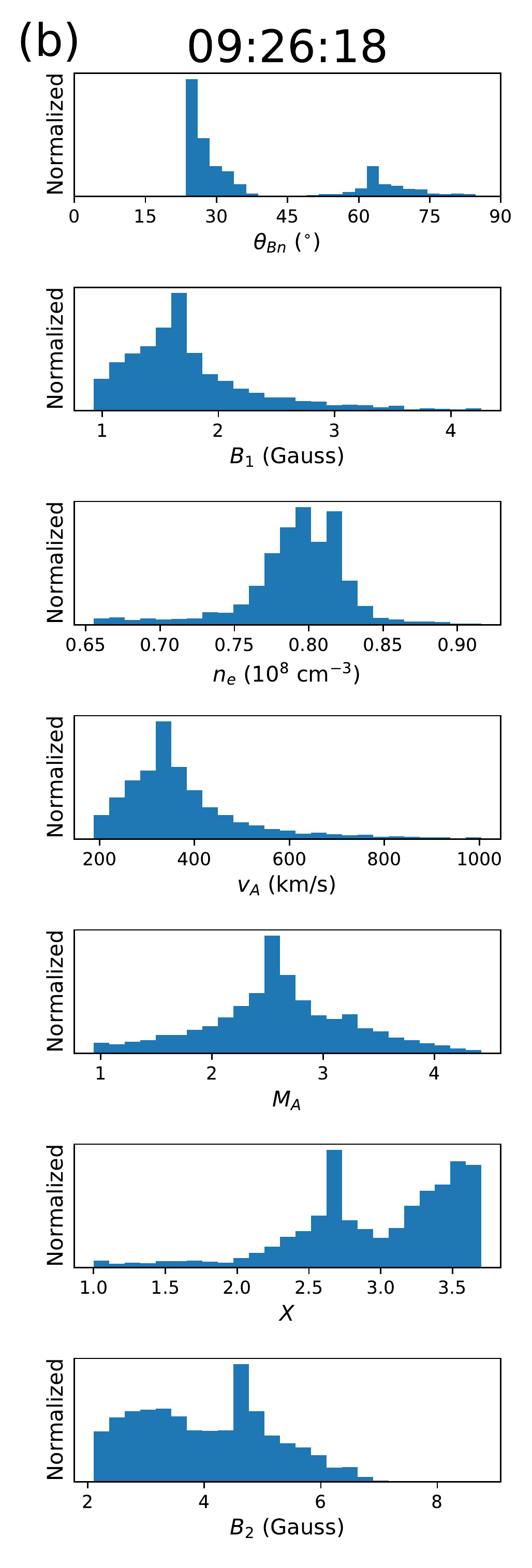}
		\end{minipage}
	\caption{Histograms of $\theta_{Bn}$, $B_1$, $n_e$, $v_A$, $M_A$, $X$, and $B_2$ of the radio source on the shock surfaces. Panels (a) and (b) show distributions of the parameters at 09:24:42 and 09:26:18 UT, respectively.}
	\label{fig:hist}
\end{figure}

\begin{figure}[ht]
	\begin{minipage}[b]{\textwidth}
		\centering
		\begin{minipage}[b]{0.12\textwidth}
			\includegraphics[width = 3.0 cm,trim = 0 3 0 0,clip = true]{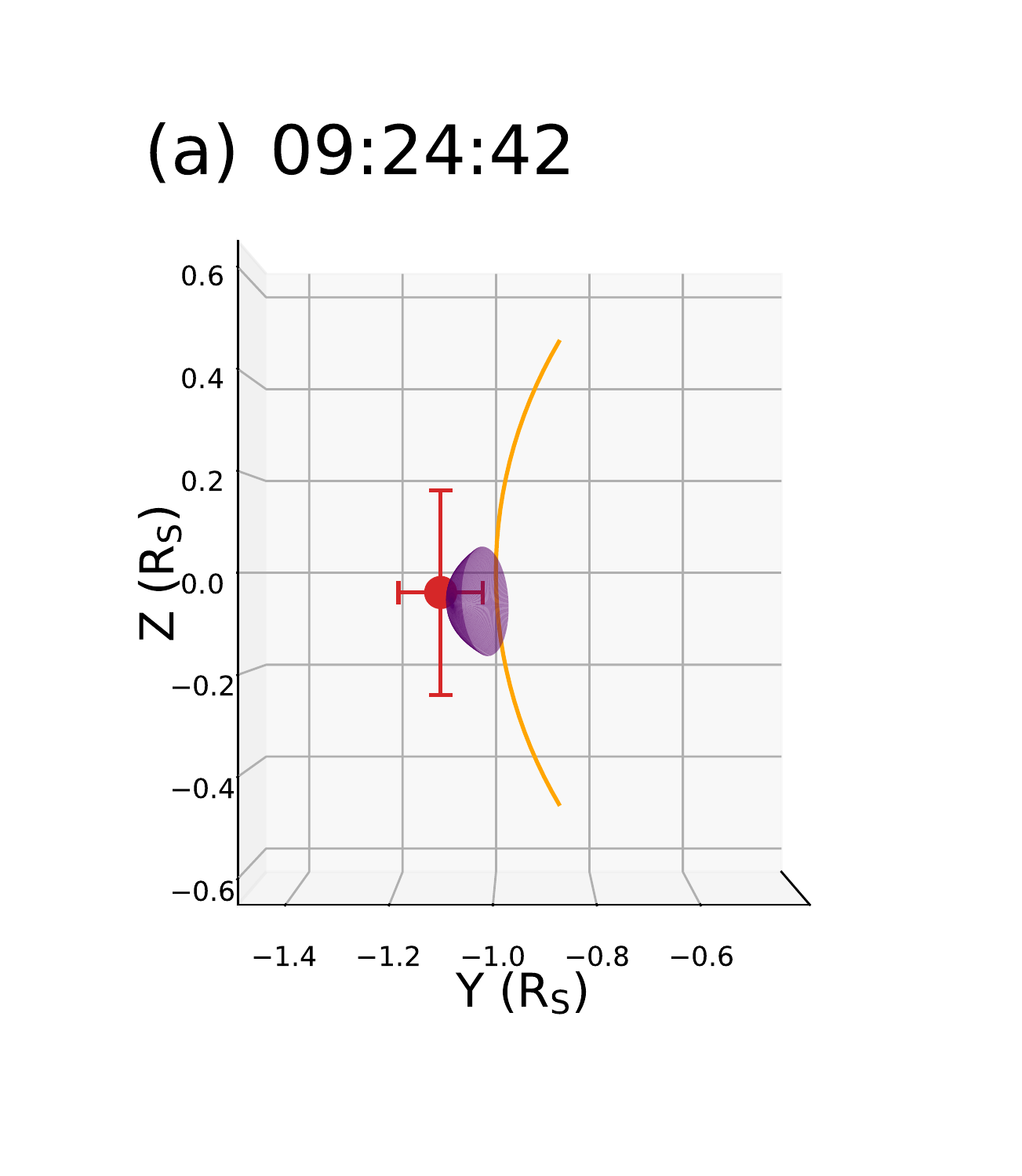}
		\end{minipage}
		\begin{minipage}[b]{0.25\textwidth}
			\centering
			\includegraphics[width = 4.11 cm,trim = 0 0 0 0,clip = true]{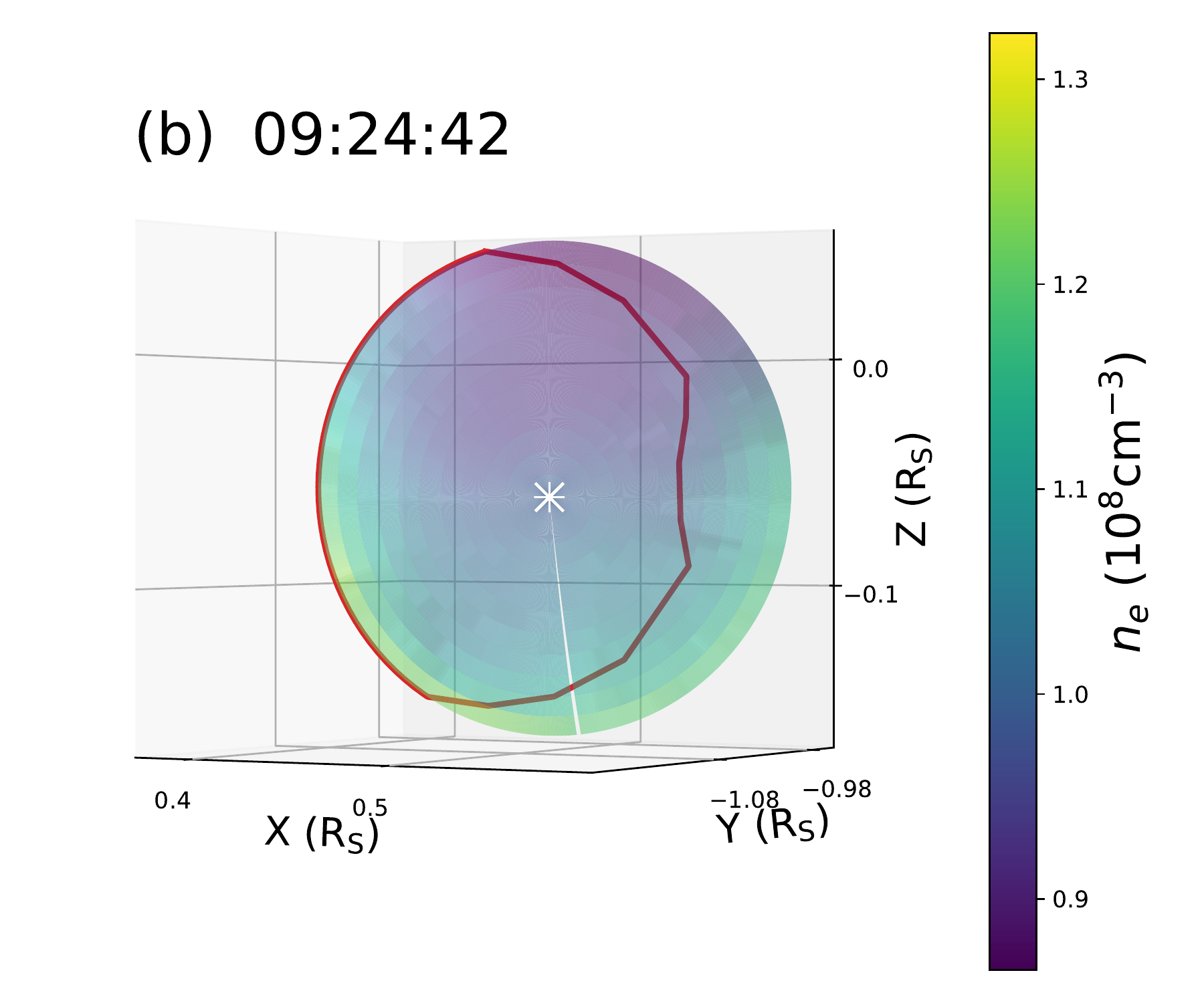}
		\end{minipage}
		\begin{minipage}[b]{0.25\textwidth}
			\centering
			\includegraphics[width = 4.11 cm,trim = 0 0 0 0,clip = true]{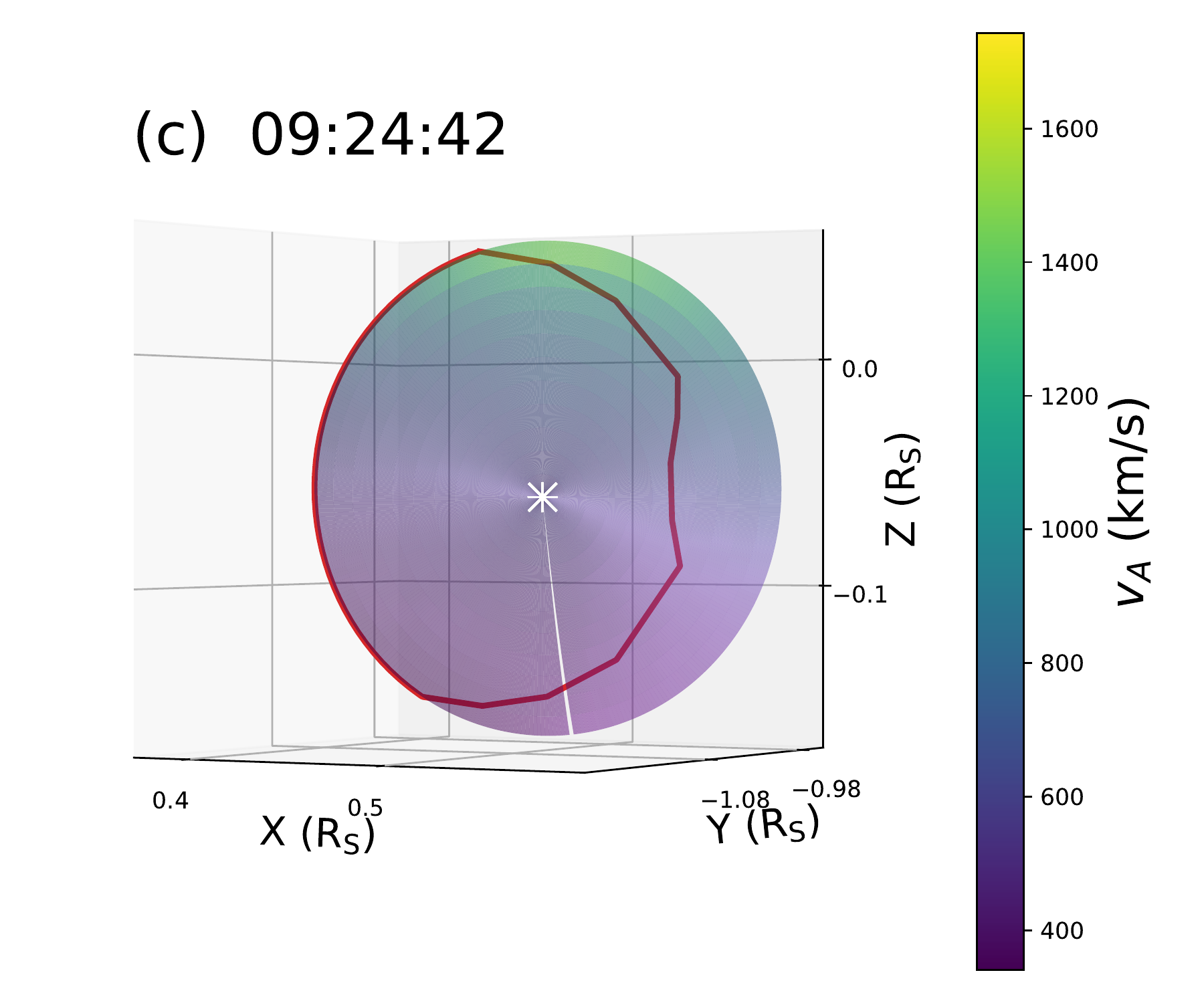}
		\end{minipage}
		\begin{minipage}[b]{0.25\textwidth}
			\centering
			\includegraphics[width = 4.11 cm,trim = 0 0 0 0,clip = true]{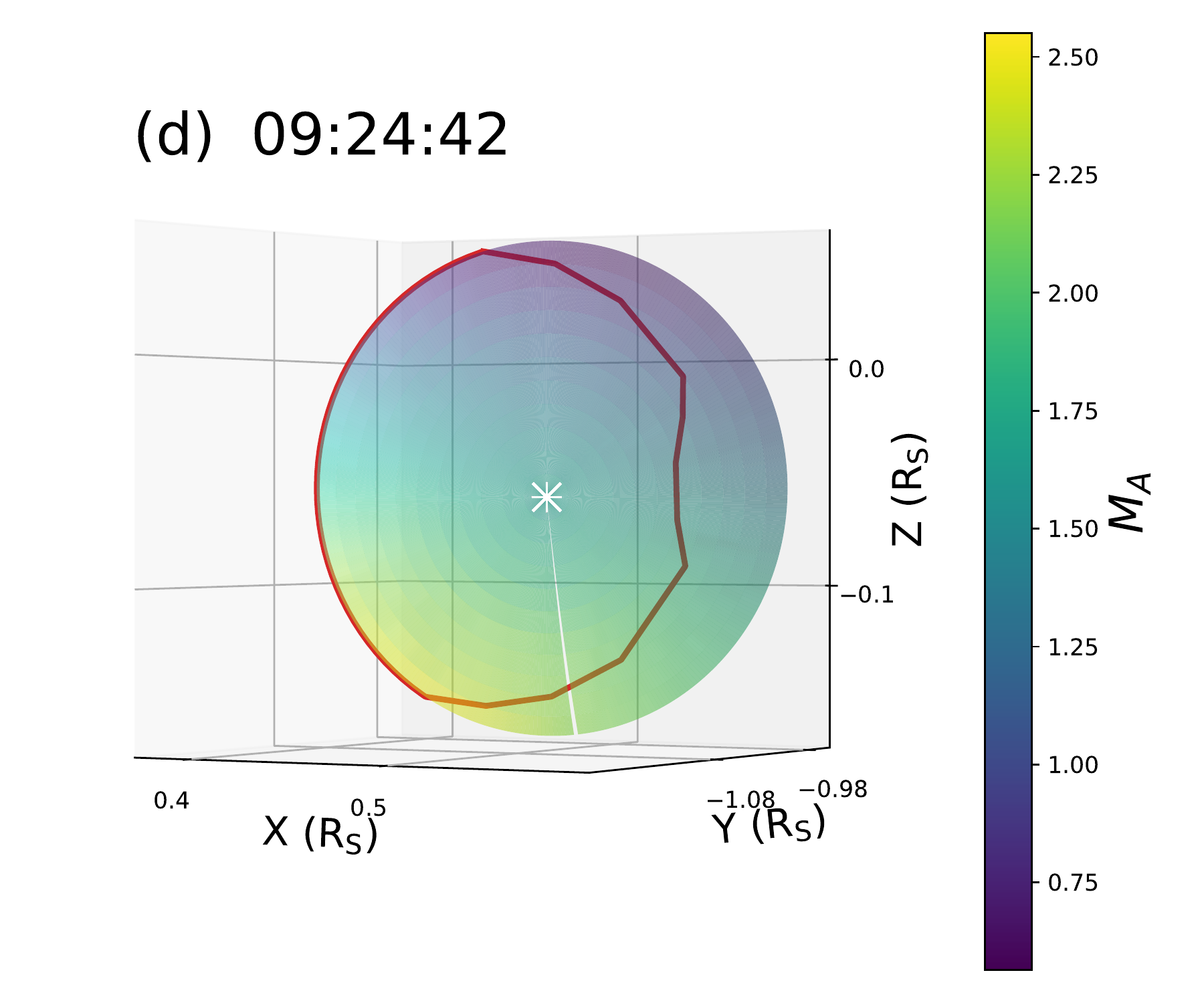}
		\end{minipage}
	\end{minipage}
	\begin{minipage}[b]{\textwidth}
		\centering
		\begin{minipage}[b]{0.12\textwidth}
			\includegraphics[width = 3 cm,trim = 0 3 0 0,clip = true]{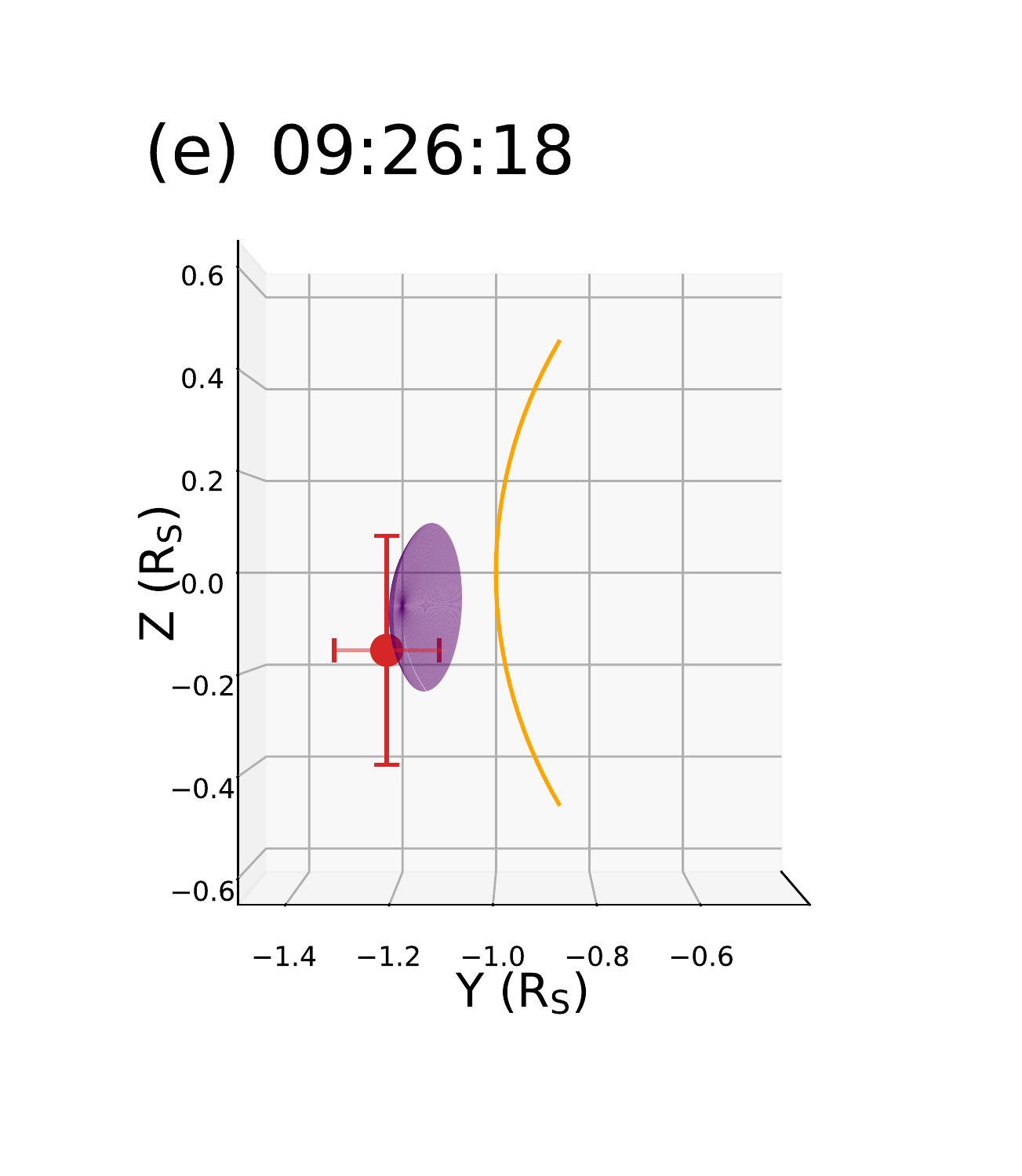}
		\end{minipage}
		\begin{minipage}[b]{0.25\textwidth}
			\centering
			\includegraphics[width = 4.11 cm,trim = 0 0 0 0,clip = true]{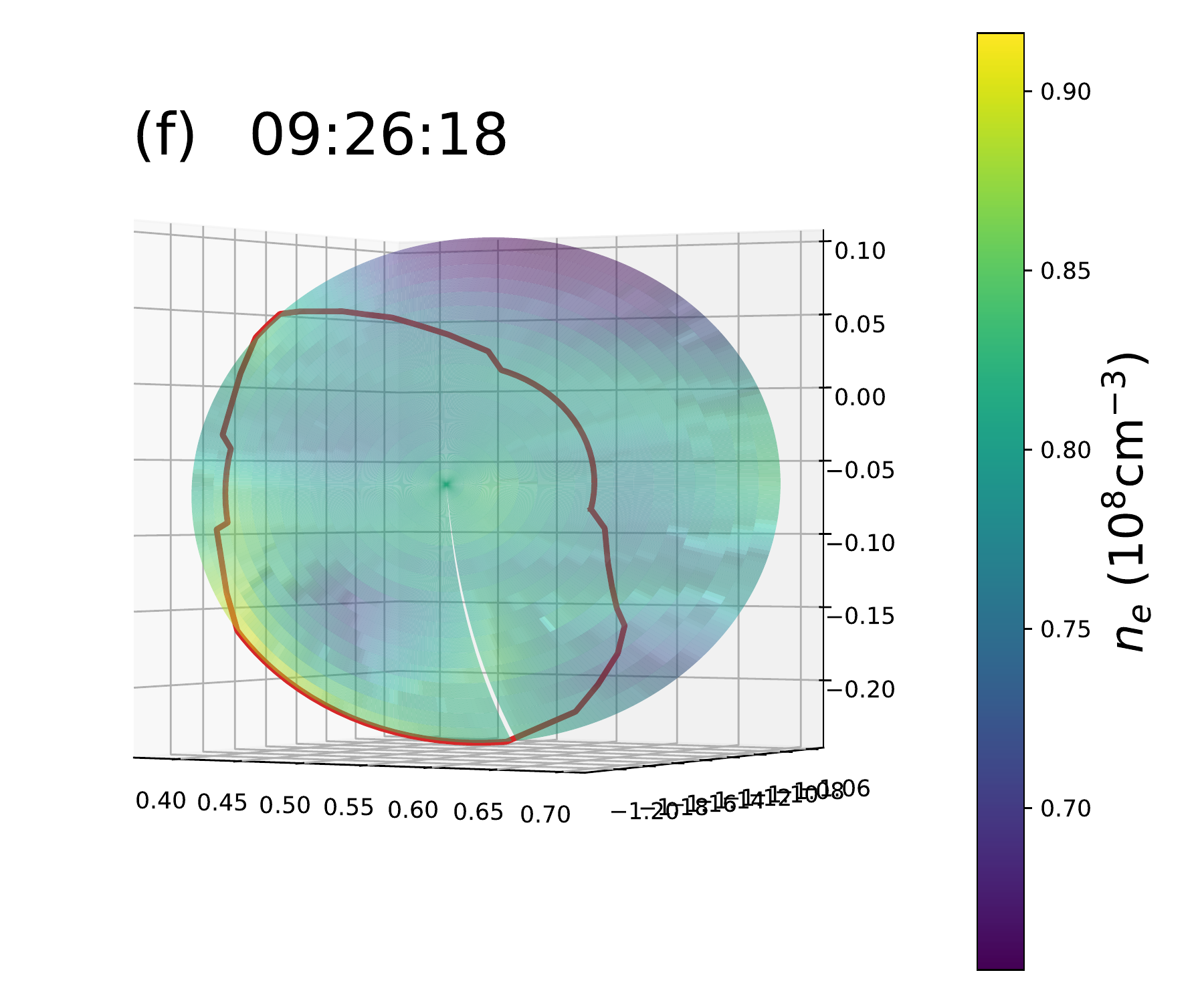}
		\end{minipage}
		\begin{minipage}[b]{0.25\textwidth}
			\centering
			\includegraphics[width = 4.11 cm,trim = 0 0 0 0,clip = true]{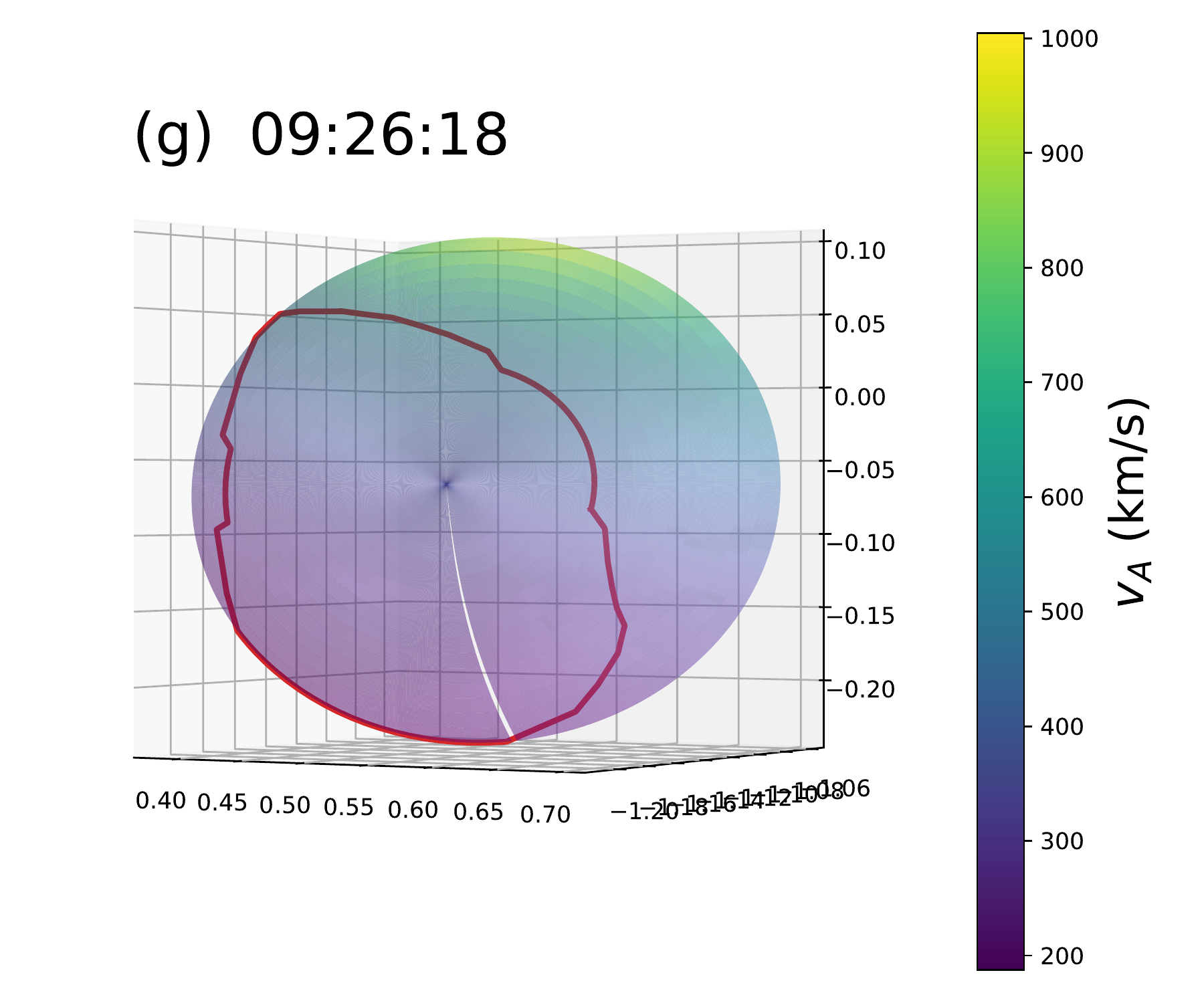}
		\end{minipage}
		\begin{minipage}[b]{0.25\textwidth}
			\centering
			\includegraphics[width = 4.11 cm,trim = 0 0 0 0,clip = true]{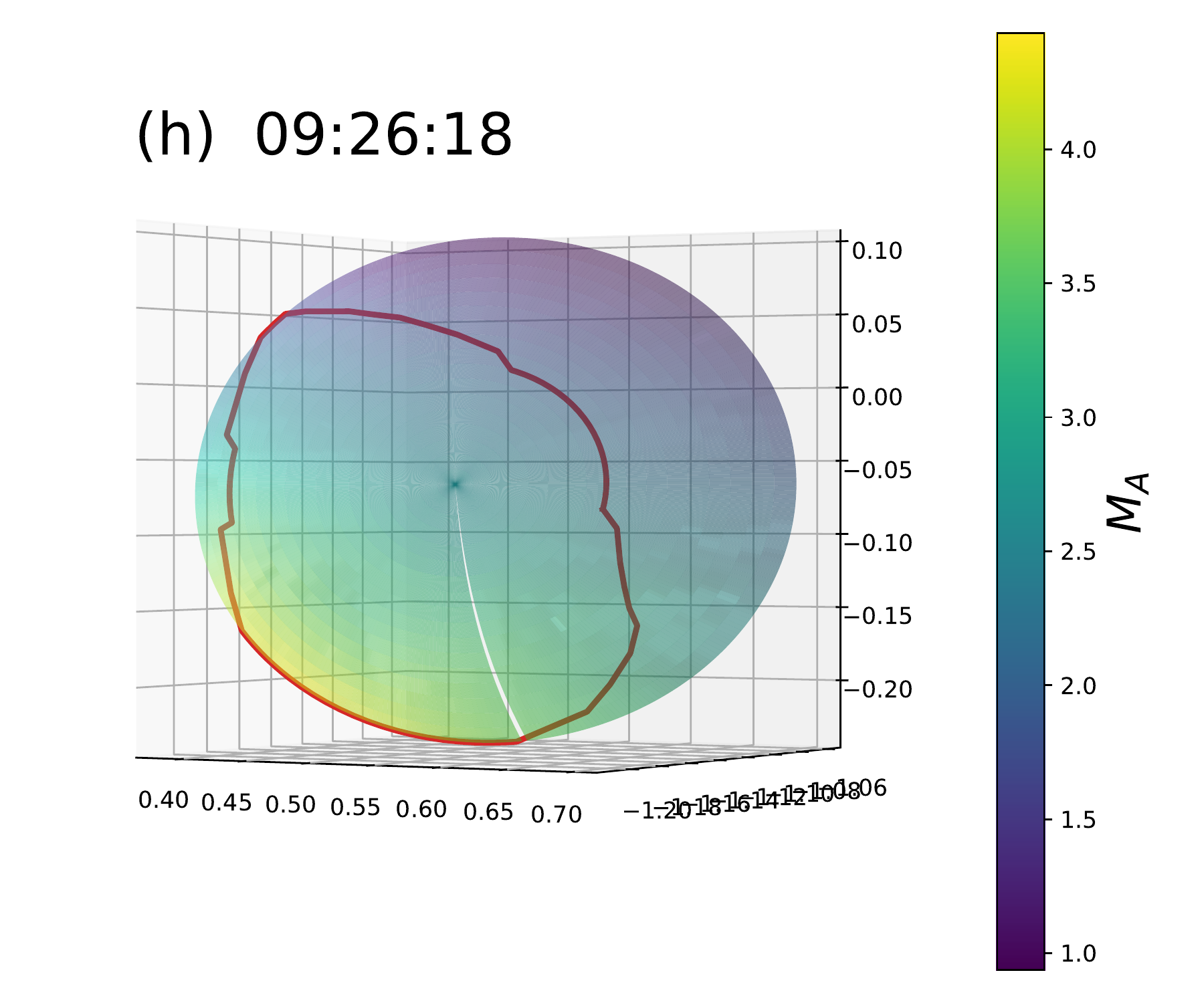}
		\end{minipage}
	\end{minipage}
	\caption{The distribution of $n_e$, $v_A$ and $M_A$ on the shock surface.
		The panels (a) and (e) show the shock surfaces with a viewing perspective from SDO in the heliocentric coordinate system at 09:24:42 and 09:26:18 UT, respectively, the red solid circles represent the radio centroids, the solar limb is represented by the orange arc.
		The upper (b, c, d) and lower (f, g, h) panels are the distribution of $n_e$, $v_A$ and $M_A$ at 09:24:42 and 09:26:18 UT, respectively. 
		The panels (b) and (f) shows the distributions of $n_e$ on the shock surfaces,
		The panels (c) and (g) shows the distributions of $v_A$ on the shock surfaces,
		the panels (d) and (h) shows the distributions of $M_A$ on the shock surfaces.
		The radio sources on the shock surface are shown as red contours in panels (b), (c), (e), and (f).
		The white asterisks are the noses of the shock surface at 09:24:42 and 09:26:18 UT.
		}
	\label{fig:MA}
\end{figure}

\begin{figure}[ht]
	\begin{minipage}[b]{\textwidth}
		\centering
		\begin{minipage}[b]{0.2\textwidth}
			\includegraphics[width = 4.5 cm,trim = 0 0 0 0,clip = true]{overview-1d.pdf}
		\end{minipage}
		\begin{minipage}[b]{0.32\textwidth}
			\includegraphics[width = 6.25 cm,trim = 0 0 0 0,clip = true]{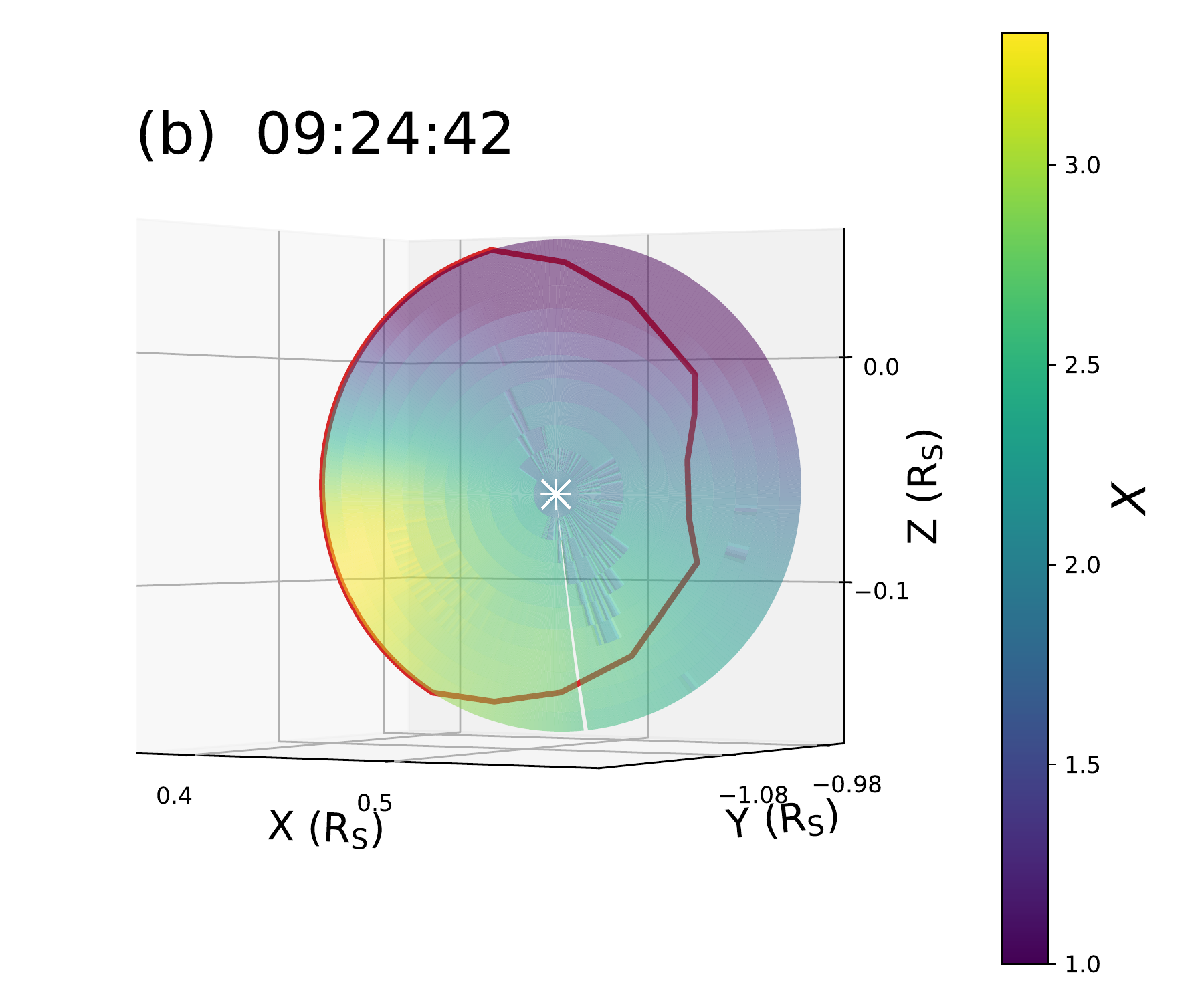}
		\end{minipage}
		\begin{minipage}[b]{0.32\textwidth}
			\includegraphics[width = 6.25 cm,trim = 0 0 0 0,clip = true]{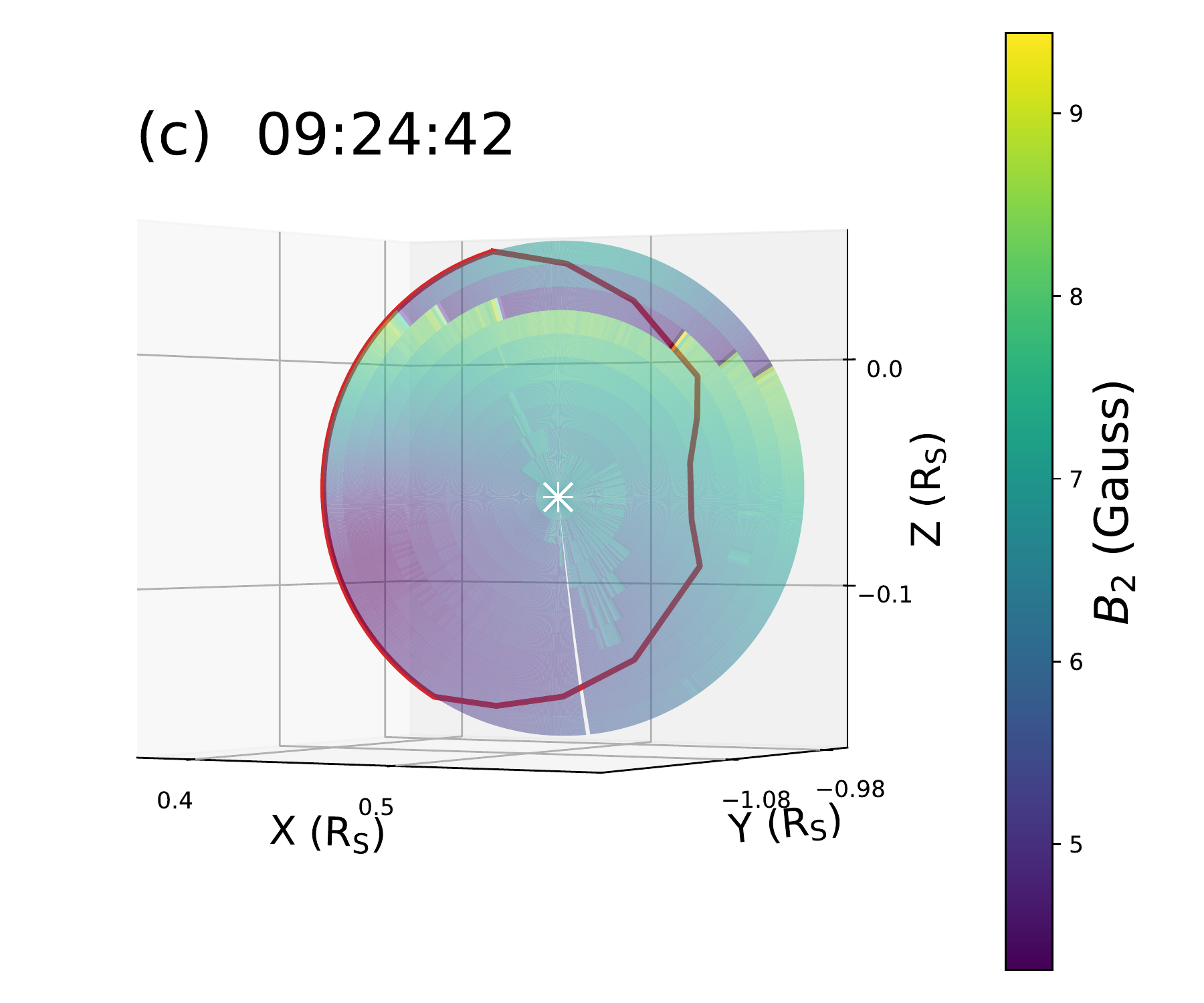}
		\end{minipage}
	\end{minipage}
	\begin{minipage}[b]{\textwidth}
		\centering
		\begin{minipage}[b]{0.2\textwidth}
			\includegraphics[width = 4.5 cm,trim = 0 0 0 0,clip = true]{overview-3d.pdf}
		\end{minipage}
		\begin{minipage}[b]{0.32\textwidth}
			\includegraphics[width = 6.25 cm,trim = 0 0 0 0,clip = true]{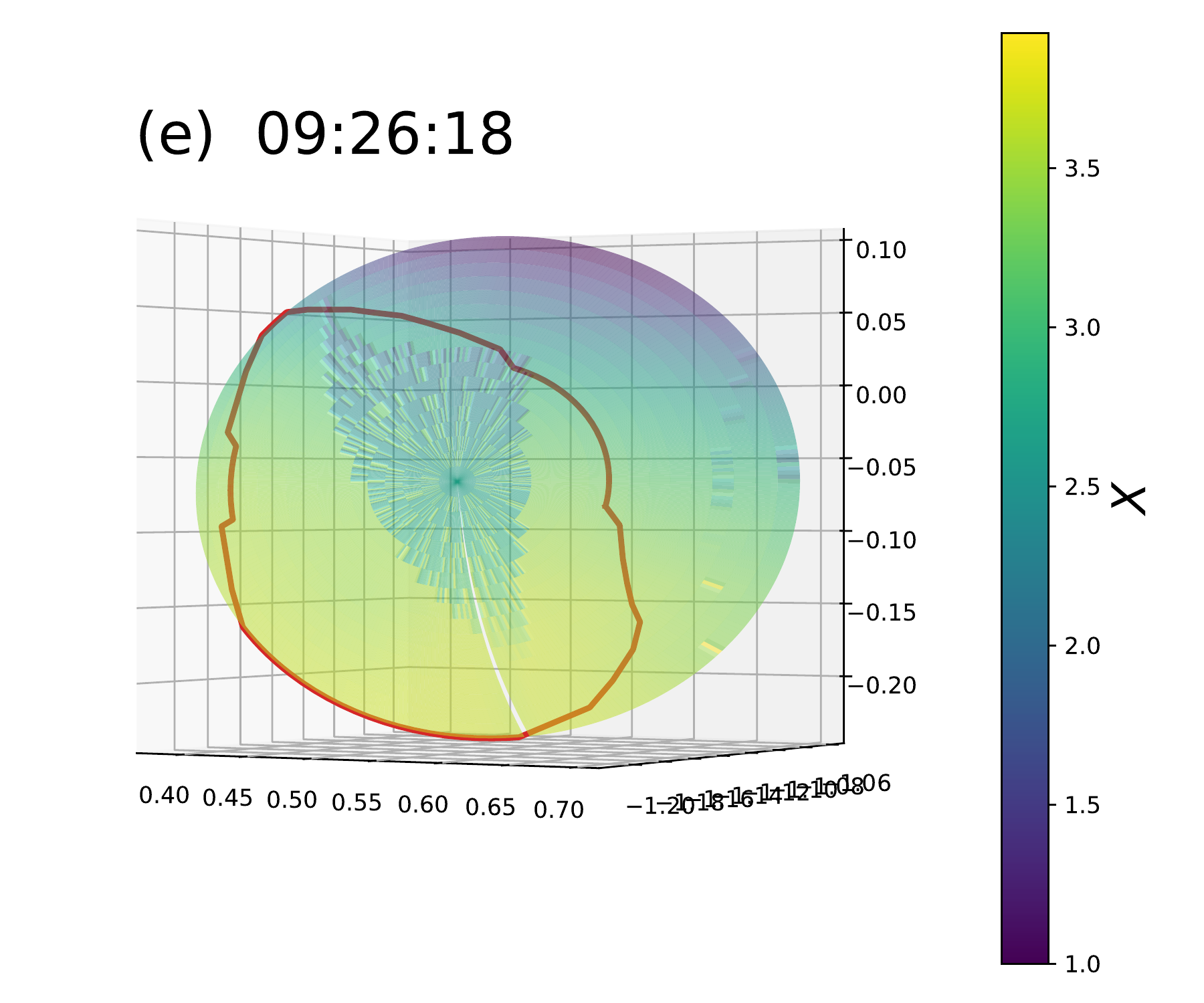}
		\end{minipage}
		\begin{minipage}[b]{0.32\textwidth}
			\includegraphics[width = 6.25 cm,trim = 0 0 0 0,clip = true]{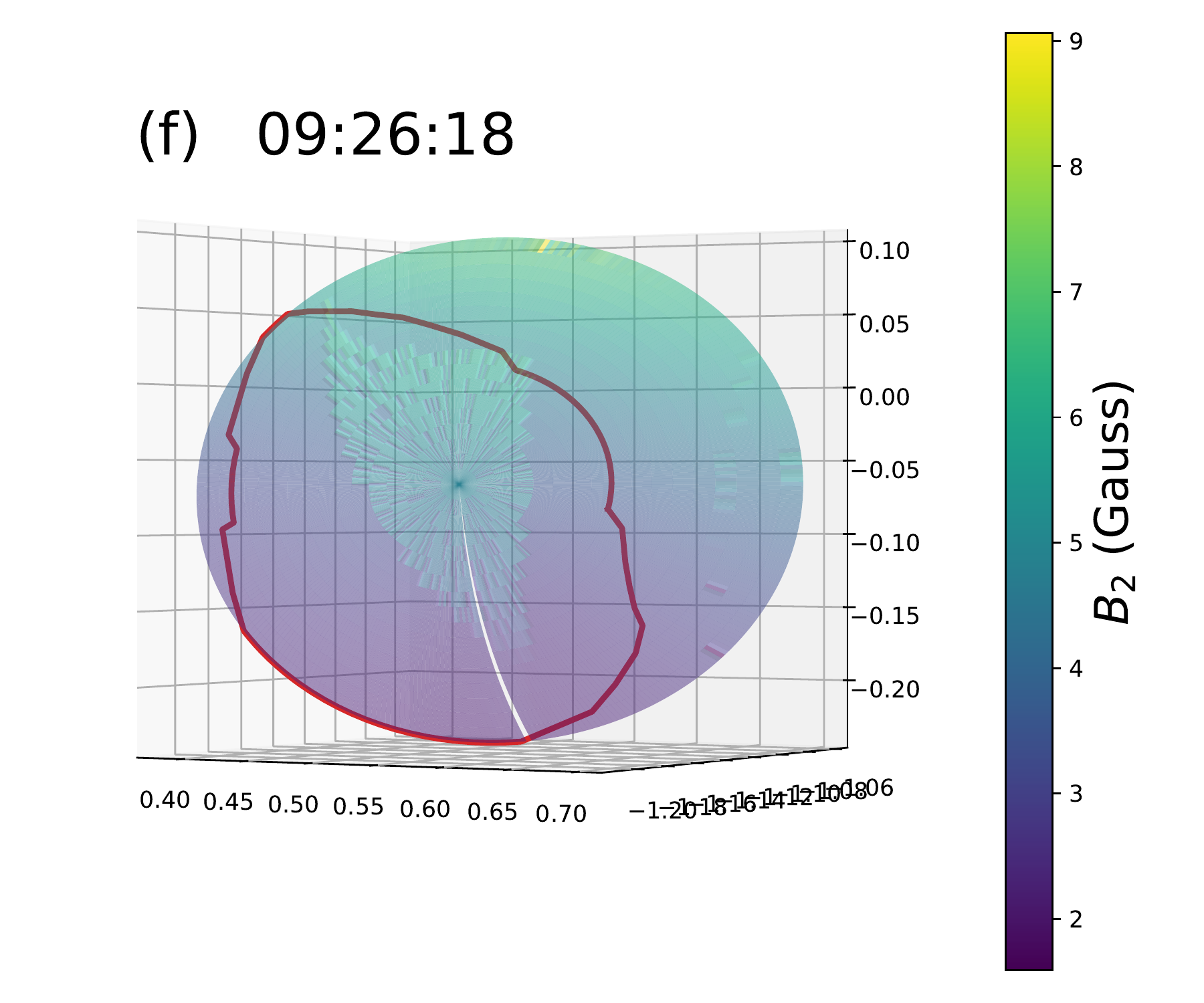}
		\end{minipage}
	\end{minipage}
	\caption{The distribution of $X$ and $B_2$ on the shock surface.
		The panels (a) and (d) show the shock surfaces with a viewing perspective from SDO in the heliocentric coordinate system at 09:24:42 and 09:26:18 UT, respectively, the red solid circles represent the radio centroids, the solar limb is represented by the orange arc.
		The upper (b, c) and lower (e, f) panels are the distribution of $X$, $B_2$ at 09:24:42 and 09:26:18 UT, respectively. 
		The panels (b) and (e) show the details of $X$ on the shock surfaces,
		the panels (c) and (f) show the details of $B_2$ on the shock surfaces.
		The radio sources on the shock surface are shown as red contours in panels (b), (c), (e), and (f).
		The white asterisks are the noses of the shock front at 09:24:42 and 09:26:18 UT.}
	\label{fig:XB2}
\end{figure}

In theory, an MHD shock can be described by the Rankine-Hugoniot (R-H) jump conditions, which result from the conservation of mass, momentum, energy, and magnetic flux \citep{Priest2014}. The plasma parameters of the shock downstream can be determined in terms of the plasma parameters of the shock upstream by the R-H relations \citep{Ruan2018}. The relation between the Alfv\'{e}n Mach number $M_A$ and the compression ratio ($X$) can be derived from the R-H relations, which is, for a perpendicular shock, $M_{\rm A\perp} = \sqrt{X(5+X)/[2(4-X)]}$, and for a parallel shock, $M_{\rm A\parallel} = \sqrt{X}$. For a general oblique shock, $M_{\rm A\angle}$ can be given under a first-order approximation as follows \citep{Bemporad2011,Bemporad2013}:
\begin{equation}
\label{equation5}
M_{\rm A\angle} = \sqrt{(M_{\rm A\perp} \rm{sin}\theta_{\rm Bn})^2 + (M_{\rm A\parallel} \rm{cos}\theta_{\rm Bn})^2}, \\
\end{equation}
where $\theta_{\rm Bn}$ is the angle between the shock normal and the upstream magnetic field vector, which has been determined in Section 3.2. Thus, combined with $\theta_{\rm Bn}$, $X$ can be derived from $M_{\rm A\angle}$. 
The distributions of $X$ on the shock surface at 09:24:42 and 09:26:18 UT are shown in Figures \ref{fig:XB2}a and \ref{fig:XB2}c.
The mean values of $X$ of the radio source on the shock surface are 2.10$\pm$0.53 and 2.99$\pm$0.45 at 09:24:42 and 09:26:18 UT, respectively.
The histograms of $X$ of the radio source on the shock surface at 09:24:42 and 09:26:18 UT are shown in Figures \ref{fig:hist}a and \ref{fig:hist}b.

From the R-H jump conditions \citep{Priest2014}, the ratio between the shock downstream and upstream magnetic field can be given as:
\begin{equation}
\label{equation6}
\left| \frac{B_2}{B_1} \right| = [\rm{cos}\theta_{\rm{Bn}}^2 + (\frac{M_{A}^2 - 1}{M_{A}^2 - X})^2 X^2 \rm{sin}\theta_{\rm{Bn}}^2]^{1/2},\\
\end{equation}
where the subscripts 1 and 2 denote the values in the up- and downstream regions, respectively. 
With the known $\theta_{\rm Bn}$, $M_{A}$, $X$, and $B_1$ from above, the magnetic field strength at shock downstream, $B_2$, can be calculated from Equation (\ref{equation6}).
The distributions of $B_2$ on the shock surface at 09:24:42 and 09:26:18 UT are shown in Figures \ref{fig:XB2}b and \ref{fig:XB2}d.
The mean values of $B_2$ of the radio source on the shock surface are 6.25$\pm$0.89 and 4.09$\pm$0.98 at 09:24:42 and 09:26:18 UT, respectively.
The histograms of $B_2$ on the shock front at 09:24:42 and 09:26:18 UT are shown in Figures \ref{fig:hist}a and \ref{fig:hist}b.

\section{Discussions} \label{results}

Coronal shocks can accelerate particles, leading to type II radio bursts. 
However, not all coronal shocks would generate type II radio bursts \citep{Gopalswamy2005,Nitta2013,Lee2014}, implying specific
properties (density, magnetic field, etc.) are needed for a coronal shock to generate type II radio bursts. 
Here, we focus on the magnetic condition for the generation of type II radio bursts. 

In the previous studies of type II radio bursts, the magnetic field of coronal shocks was sometimes assumed to be simple in geometry \citep{Ma2011,Su2015,Susino2015}. Or the magnetic field structure at the radio source region was inferred when type II radio bursts are excited by coronal shocks passing through some specific structures, such as coronal streamers \citep{Kong2012} or CME/flare current sheets \citep{Gao2016}. These approximations may not be accurate enough. Moreover, it is not expected that type II radio bursts are all generated at these structures. 
Even though the average magnetic field strength of coronal shocks can be estimated from the radio dynamic spectra \citep{Mann1995a,Vrsnak2002}, this approach cannot get the orientation and spatial distribution of the magnetic field.
In this work, we use the coronal magnetic field extrapolation to obtain the 3D distribution of the magnetic field on the shock front. 
Combined with the fitted shock front, we can obtain the shock geometry.

We obtained the distribtions of $\theta_{\rm Bn}$, $B_1$, $v_A$, and $M_A$ on the shock front (Figures \ref{fig:thetaOnShock} and \ref{fig:MA}).
Due to the location uncertainty of the radio source, the statistical properties of the distributions of the radio source are used to characterize the physical conditions of the type II burst generation.
We found that the distributions of $B_1$, $v_A$, and $M_A$ at the shock front are roughly unimodal with some asymmetry, but the distribution of $\theta_{\rm Bn}$ is bimodal (Figure \ref{fig:hist}).
The parameters of the shock downstream can be derived by the R-H relations as long as the parameters of the shock upstream are given. The R-H relations have been widely applied to the in-situ observations in the interplanetary space \citep{Wang2018}. 
For the coronal shock we are investigating in, only remote observations are available. We cannot get the physical parameters of the shock directly. 
Therefore, without an assumption of a quasi-perpendicular or quasi-parallel shock geometry, 
we calculate the values of $\theta_{\rm Bn}$ of the shock from the extrapolated PFSS magnetic fields on the 3D shock surface.
There values are then applied to derive $M_A$, $X$, and the magnetic field strength of the shock downstream, $B_2$ by using the R-H relations. 

In this work, the mean value of $M_A$ of the shock front, a parameter describing the strength of the shock, is about 2 for the current event, which is quantitatively consistent with the previous results \citep{Bemporad2011,Bemporad2014,Su2016}. 
Meanwhile, those studies all indicated that type II radio bursts are generated at the region where $M_A$ is the largest. 
This finding implies that there is possibly a threshold of $M_A$ in the generation of type II radio bursts, which is about 1.5 as revealed in this work.

Energetic electrons are necessary for the generation of type II radio bursts. Both quasi-perpendicular and quasi-parallel shocks can accelerate electrons effectively \citep{Mann2001,Masters2013,Qin2018,Kong2020}.
For some events, it was identified that the type II radio bursts are excited by quasi-perpendicular shocks \citep{Zucca2018,Maguire2020}.
Combining the histogram (Figure \ref{fig:hist}) and statistical results (Table \ref{table1}) of $\theta_{Bn}$, it show that the type II radio burst is generated by a shock with a bimodal distribution of $\theta_{Bn}$ and average value of $\theta_{Bn} \approx 40^{\circ}$, it implies that both quasi-perpendicular and quasi-parallel shock structures may be at work for emitting the type II radio burst.
Owing to the uncertainty of the radio source location and the spatial resolution of the NRH, we cannot distinguish the role played by quasi-perpendicular and quasi-parallel structures in this study. 
In order to reveal the role played by the fine structures of coronal shocks in the generation of type II bursts, a higher spatial resolution of radio observation and/or new technologies are needed in the future, e.g., using the dispersion effect, the multi-frequencies scheme might be effective for suppressing the uncertainty of the radio location.

\section{Conclusions} \label{results}

In summary, we explored the 3D magnetic conditions for the generation of a type II radio burst. To this end
we first reconstructed the 3D shock surface based on the multi-perspective EUV observations by SDO and STEREO. 
We then used the PFSS model to quantify the distribtutions of $\theta_{\rm Bn}$ and $B_1$, 
and furthermore the distributions of $v_A$ and $M_A$ on the 3D shock front. 
Combined with the radio observation of the NRH, we found that the type II radio burst is generated by a shock with $M_A \gtrsim 1.5$ and a bimodal distribution of $\theta_{Bn}$. Besides, we used the R-H jump relations to obtain the distributions of the shock downstream properties ($X$ and $B_2$) on the 3D shock front.

Combined with new proposals of multi-scale in-situ observations \citep{Retino2019,Dai2020} and magnetic field extrapolation models, more details of the bimodal distribution of $\theta_{\rm Bn}$ are expected to be revealed from MHD scales to plasma scales.
Furthermore, since the supernova remnants can also be revealed in the radio images and can be identified as quasi-perpendicular or quasi-parallel \citep{Reynoso2013}, 
our results can be compared in the future with other MHD shocks in astrophysics, such as the supernova remnants.

\acknowledgments
We thank SunPy Community for conveniently plotting in this work \citep{SunPy2015}. We thank the SDO team for providing the EUV images, the STEREO team for 3D observations, and C. Schrijver and M. DeRosa for providing PFSS code. We are grateful to Zucca P., Dai Y., Sun S.D., Chen Y. and Zhao G.Q. for valuable discussions. 
S.W. is supported by National Key R \& D Program of China (Grant 2020YFC2201200), the National Natural Science Foundation of China (NSFC) under grants 11803008, 11773079, 91636111, 11690021, 11973024, 11773016, 11733003, 11961131002, 11533005, Jiangsu NSF (BK20171108).

\bibliography{sample63}{}

\begin{thebibliography}{}
\expandafter\ifx\csname natexlab\endcsname\relax\def\natexlab#1{#1}\fi
\providecommand{\url}[1]{\href{#1}{#1}}
\providecommand{\dodoi}[1]{doi:~\href{http://doi.org/#1}{\nolinkurl{#1}}}
\providecommand{\doeprint}[1]{\href{http://ascl.net/#1}{\nolinkurl{http://ascl.net/#1}}}
\providecommand{\doarXiv}[1]{\href{https://arxiv.org/abs/#1}{\nolinkurl{https://arxiv.org/abs/#1}}}

\bibitem[{{Aschwanden} {et~al.}(2001){Aschwanden}, {Schrijver}, \&
  {Alexander}}]{Aschwanden2001}
{Aschwanden}, M.~J., {Schrijver}, C.~J., \& {Alexander}, D. 2001, \apj, 550,
  1036, \dodoi{10.1086/319796}

\bibitem[{{Bemporad} \& {Mancuso}(2010)}]{Bemporad2010}
{Bemporad}, A., \& {Mancuso}, S. 2010, \apj, 720, 130,
  \dodoi{10.1088/0004-637X/720/1/130}

\bibitem[{{Bemporad} \& {Mancuso}(2011)}]{Bemporad2011}
---. 2011, \apjl, 739, L64, \dodoi{10.1088/2041-8205/739/2/L64}

\bibitem[{{Bemporad} \& {Mancuso}(2013)}]{Bemporad2013}
---. 2013, Journal of Advanced Research, 4, 287,
  \dodoi{10.1016/j.jare.2012.09.005}

\bibitem[{{Bemporad} {et~al.}(2014){Bemporad}, {Susino}, \&
  {Lapenta}}]{Bemporad2014}
{Bemporad}, A., {Susino}, R., \& {Lapenta}, G. 2014, \apj, 784, 102,
  \dodoi{10.1088/0004-637X/784/2/102}

\bibitem[{{Brueckner} {et~al.}(1995){Brueckner}, {Howard}, {Koomen},
  {Korendyke}, {Michels}, {Moses}, {Socker}, {Dere}, {Lamy}, {Llebaria},
  {Bout}, {Schwenn}, {Simnett}, {Bedford}, \& {Eyles}}]{Brueckner1995}
{Brueckner}, G.~E., {Howard}, R.~A., {Koomen}, M.~J., {et~al.} 1995, \solphys,
  162, 357, \dodoi{10.1007/BF00733434}

\bibitem[{{Cairns}(1986)}]{Cairns1986}
{Cairns}, I.~H. 1986, \jgr, 91, 2975, \dodoi{10.1029/JA091iA03p02975}

\bibitem[{{Cairns}(1988)}]{Cairns1988}
---. 1988, \jgr, 93, 858, \dodoi{10.1029/JA093iA02p00858}

\bibitem[{{Cairns} \& {Melrose}(1985)}]{Cairns1985}
{Cairns}, I.~H., \& {Melrose}, D.~B. 1985, \jgr, 90, 6637,
  \dodoi{10.1029/JA090iA07p06637}

\bibitem[{{Chen} {et~al.}(2015){Chen}, {Bastian}, {Shen}, {Gary}, {Krucker}, \&
  {Glesener}}]{ChenB2015}
{Chen}, B., {Bastian}, T.~S., {Shen}, C., {et~al.} 2015, Science, 350, 1238,
  \dodoi{10.1126/science.aac8467}

\bibitem[{{Chen}(2011)}]{Chen2011}
{Chen}, P.~F. 2011, Living Reviews in Solar Physics, 8, 1,
  \dodoi{10.12942/lrsp-2011-1}

\bibitem[{{Chen} {et~al.}(2020){Chen}, {Kontar}, {Chrysaphi}, {Jeffrey},
  {Gordovskyy}, {Yan}, \& {Tan}}]{Chen2020}
{Chen}, X., {Kontar}, E.~P., {Chrysaphi}, N., {et~al.} 2020, \apj, 905, 43,
  \dodoi{10.3847/1538-4357/abc24e}

\bibitem[{{Chen} {et~al.}(2014){Chen}, {Du}, {Feng}, {Feng}, {Kong}, {Guo},
  {Wang}, \& {Li}}]{Chen2014}
{Chen}, Y., {Du}, G., {Feng}, L., {et~al.} 2014, \apj, 787, 59,
  \dodoi{10.1088/0004-637X/787/1/59}

\bibitem[{{Cheng} {et~al.}(2012){Cheng}, {Zhang}, {Saar}, \&
  {Ding}}]{Cheng2012}
{Cheng}, X., {Zhang}, J., {Saar}, S.~H., \& {Ding}, M.~D. 2012, \apj, 761, 62,
  \dodoi{10.1088/0004-637X/761/1/62}

\bibitem[{Dai {et~al.}(2020)Dai, Wang, Cai, Gonzalez, Hesse, Escoubet, Phan,
  Vasyliunas, Lu, Li, Kong, Dunlop, Nakamura, He, Fu, Zhou, Huang, Wang,
  Khotyaintsev, Graham, Retino, Zelenyi, Grigorenko, Runov, Angelopoulos,
  Kepko, Hwang, \& Zhang}]{Dai2020}
Dai, L., Wang, C., Cai, Z., {et~al.} 2020, Frontiers in Physics, 8, 89,
  \dodoi{10.3389/fphy.2020.00089}

\bibitem[{{Draine} \& {McKee}(1993)}]{Draine1993}
{Draine}, B.~T., \& {McKee}, C.~F. 1993, \araa, 31, 373,
  \dodoi{10.1146/annurev.aa.31.090193.002105}

\bibitem[{{Du} {et~al.}(2014){Du}, {Chen}, {Lv}, {Kong}, {Feng}, {Guo}, \&
  {Li}}]{Du2014}
{Du}, G., {Chen}, Y., {Lv}, M., {et~al.} 2014, \apjl, 793, L39,
  \dodoi{10.1088/2041-8205/793/2/L39}

\bibitem[{{Feng} {et~al.}(2020){Feng}, {Lu}, {Inhester}, {Plowman}, {Ying},
  {Mierla}, {West}, \& {Gan}}]{Feng2020}
{Feng}, L., {Lu}, L., {Inhester}, B., {et~al.} 2020, \solphys, 295, 141,
  \dodoi{10.1007/s11207-020-01710-3}

\bibitem[{{Frassati} {et~al.}(2019){Frassati}, {Susino}, {Mancuso}, \&
  {Bemporad}}]{Frassati2019}
{Frassati}, F., {Susino}, R., {Mancuso}, S., \& {Bemporad}, A. 2019, \apj, 871,
  212, \dodoi{10.3847/1538-4357/aaf9af}

\bibitem[{{Gao} {et~al.}(2016){Gao}, {Wang}, {Wu}, {Lin}, {Ebenezer}, \&
  {Tan}}]{Gao2016}
{Gao}, G., {Wang}, M., {Wu}, N., {et~al.} 2016, \solphys, 291, 3369,
  \dodoi{10.1007/s11207-016-1007-x}

\bibitem[{{Ginzburg} \& {Zhelezniakov}(1958)}]{Ginzburg1958}
{Ginzburg}, V.~L., \& {Zhelezniakov}, V.~V. 1958, \sovast, 2, 653

\bibitem[{{Gopalswamy} {et~al.}(2005){Gopalswamy}, {Aguilar-Rodriguez},
  {Yashiro}, {Nunes}, {Kaiser}, \& {Howard}}]{Gopalswamy2005}
{Gopalswamy}, N., {Aguilar-Rodriguez}, E., {Yashiro}, S., {et~al.} 2005,
  Journal of Geophysical Research (Space Physics), 110, 12,
  \dodoi{10.1029/2005JA011158}

\bibitem[{{Gopalswamy} {et~al.}(2009){Gopalswamy}, {Thompson}, {Davila},
  {Kaiser}, {Yashiro}, {M{\"a}kel{\"a}}, {Michalek}, {Bougeret}, \&
  {Howard}}]{Gopalswamy2009}
{Gopalswamy}, N., {Thompson}, W.~T., {Davila}, J.~M., {et~al.} 2009, \solphys,
  259, 227, \dodoi{10.1007/s11207-009-9382-1}

\bibitem[{{Hao} {et~al.}(2017){Hao}, {Gao}, {Lu}, {Huang}, {Wang}, \&
  {Wang}}]{Hao2017}
{Hao}, Y., {Gao}, X., {Lu}, Q., {et~al.} 2017, Journal of Geophysical Research
  (Space Physics), 122, 6385, \dodoi{10.1002/2017JA024234}

\bibitem[{{Kaiser} {et~al.}(2008){Kaiser}, {Kucera}, {Davila}, {St.~Cyr},
  {Guhathakurta}, \& {Christian}}]{Kaiser2008}
{Kaiser}, M.~L., {Kucera}, T.~A., {Davila}, J.~M., {et~al.} 2008, \ssr, 136, 5,
  \dodoi{10.1007/s11214-007-9277-0}

\bibitem[{{Kerdraon} \& {Delouis}(1997)}]{Kerdraon1997}
{Kerdraon}, A., \& {Delouis}, J.-M. 1997, in Lecture Notes in Physics, Berlin
  Springer Verlag, Vol. 483, Coronal Physics from Radio and Space Observations,
  ed. G.~{Trottet}, 192, \dodoi{10.1007/BFb0106458}

\bibitem[{{Knock} \& {Cairns}(2005)}]{Knock2005}
{Knock}, S.~A., \& {Cairns}, I.~H. 2005, Journal of Geophysical Research (Space
  Physics), 110, A01101, \dodoi{10.1029/2004JA010452}

\bibitem[{{Kong} \& {Qin}(2020)}]{Kong2020}
{Kong}, F.~J., \& {Qin}, G. 2020, \apj, 896, 20,
  \dodoi{10.3847/1538-4357/ab8e32}

\bibitem[{{Kong} {et~al.}(2012){Kong}, {Chen}, {Li}, {Feng}, {Song}, {Guo}, \&
  {Jiao}}]{Kong2012}
{Kong}, X.~L., {Chen}, Y., {Li}, G., {et~al.} 2012, \apj, 750, 158,
  \dodoi{10.1088/0004-637X/750/2/158}

\bibitem[{{Kontar} {et~al.}(2017){Kontar}, {Yu}, {Kuznetsov}, {Emslie},
  {Alcock}, {Jeffrey}, {Melnik}, {Bian}, \& {Subramanian}}]{Kontar2017}
{Kontar}, E.~P., {Yu}, S., {Kuznetsov}, A.~A., {et~al.} 2017, Nature
  Communications, 8, 1515, \dodoi{10.1038/s41467-017-01307-8}

\bibitem[{{Lee} {et~al.}(2014){Lee}, {Moon}, {Lee}, {Lee}, {Kim}, \&
  {Lee}}]{Lee2014}
{Lee}, J.-O., {Moon}, Y.-J., {Lee}, J.-Y., {et~al.} 2014, \apjl, 796, L16,
  \dodoi{10.1088/2041-8205/796/1/L16}

\bibitem[{{Lemen} {et~al.}(2012){Lemen}, {Title}, {Akin}, {Boerner}, {Chou},
  {Drake}, {Duncan}, {Edwards}, {Friedlaender}, {Heyman}, {Hurlburt}, {Katz},
  {Kushner}, {Levay}, {Lindgren}, {Mathur}, {McFeaters}, {Mitchell}, {Rehse},
  {Schrijver}, {Springer}, {Stern}, {Tarbell}, {Wuelser}, {Wolfson}, {Yanari},
  {Bookbinder}, {Cheimets}, {Caldwell}, {Deluca}, {Gates}, {Golub}, {Park},
  {Podgorski}, {Bush}, {Scherrer}, {Gummin}, {Smith}, {Auker}, {Jerram},
  {Pool}, {Soufli}, {Windt}, {Beardsley}, {Clapp}, {Lang}, \&
  {Waltham}}]{Lemen2012}
{Lemen}, J.~R., {Title}, A.~M., {Akin}, D.~J., {et~al.} 2012, \solphys, 275,
  17, \dodoi{10.1007/s11207-011-9776-8}

\bibitem[{{Lu} {et~al.}(2017){Lu}, {Inhester}, {Feng}, {Liu}, \&
  {Zhao}}]{Lu2017}
{Lu}, L., {Inhester}, B., {Feng}, L., {Liu}, S., \& {Zhao}, X. 2017, \apj, 835,
  188, \dodoi{10.3847/1538-4357/835/2/188}

\bibitem[{{Ma} {et~al.}(2011){Ma}, {Raymond}, {Golub}, {Lin}, {Chen}, {Grigis},
  {Testa}, \& {Long}}]{Ma2011}
{Ma}, S., {Raymond}, J.~C., {Golub}, L., {et~al.} 2011, \apj, 738, 160,
  \dodoi{10.1088/0004-637X/738/2/160}

\bibitem[{{Magdaleni{\'c}} {et~al.}(2012){Magdaleni{\'c}}, {Marqu{\'e}},
  {Zhukov}, {Vr{\v s}nak}, \& {Veronig}}]{Magdalenic2012}
{Magdaleni{\'c}}, J., {Marqu{\'e}}, C., {Zhukov}, A.~N., {Vr{\v s}nak}, B., \&
  {Veronig}, A. 2012, \apj, 746, 152, \dodoi{10.1088/0004-637X/746/2/152}

\bibitem[{{Maguire} {et~al.}(2020){Maguire}, {Carley}, {McCauley}, \&
  {Gallagher}}]{Maguire2020}
{Maguire}, C.~A., {Carley}, E.~P., {McCauley}, J., \& {Gallagher}, P.~T. 2020,
  \aap, 633, A56, \dodoi{10.1051/0004-6361/201936449}

\bibitem[{{Mann} \& {Classen}(1995)}]{Mann1995b}
{Mann}, G., \& {Classen}, H.-T. 1995, \aap, 304, 576

\bibitem[{{Mann} {et~al.}(2001){Mann}, {Classen}, \& {Motschmann}}]{Mann2001}
{Mann}, G., {Classen}, H.~T., \& {Motschmann}, U. 2001, \jgr, 106, 25323,
  \dodoi{10.1029/2000JA004010}

\bibitem[{{Mann} {et~al.}(1995){Mann}, {Classen}, \& {Aurass}}]{Mann1995a}
{Mann}, G., {Classen}, T., \& {Aurass}, H. 1995, \aap, 295, 775

\bibitem[{{Mann} {et~al.}(1999){Mann}, {Jansen}, {MacDowall}, {Kaiser}, \&
  {Stone}}]{Mann1999}
{Mann}, G., {Jansen}, F., {MacDowall}, R.~J., {Kaiser}, M.~L., \& {Stone},
  R.~G. 1999, \aap, 348, 614

\bibitem[{{Mann} \& {Klassen}(2005)}]{Mann2005}
{Mann}, G., \& {Klassen}, A. 2005, \aap, 441, 319,
  \dodoi{10.1051/0004-6361:20034396}

\bibitem[{{Mann} {et~al.}(2018){Mann}, {Melnik}, {Rucker}, {Konovalenko}, \&
  {Brazhenko}}]{Mann2018}
{Mann}, G., {Melnik}, V.~N., {Rucker}, H.~O., {Konovalenko}, A.~A., \&
  {Brazhenko}, A.~I. 2018, \aap, 609, A41, \dodoi{10.1051/0004-6361/201730546}

\bibitem[{{Masters} {et~al.}(2013){Masters}, {Stawarz}, {Fujimoto}, {Schwartz},
  {Sergis}, {Thomsen}, {Retin{\`o}}, {Hasegawa}, {Zieger}, {Lewis}, {Coates},
  {Canu}, \& {Dougherty}}]{Masters2013}
{Masters}, A., {Stawarz}, L., {Fujimoto}, M., {et~al.} 2013, Nature Physics, 9,
  164, \dodoi{10.1038/nphys2541}

\bibitem[{{Morosan} {et~al.}(2020){Morosan}, {Palmerio}, {Pomoell}, {Vainio},
  {Palmroth}, \& {Kilpua}}]{Morosan2020}
{Morosan}, D.~E., {Palmerio}, E., {Pomoell}, J., {et~al.} 2020, \aap, 635, A62,
  \dodoi{10.1051/0004-6361/201937133}

\bibitem[{{Morosan} {et~al.}(2019){Morosan}, {Carley}, {Hayes}, {Murray},
  {Zucca}, {Fallows}, {McCauley}, {Kilpua}, {Mann}, {Vocks}, \&
  {Gallagher}}]{Morosan2019}
{Morosan}, D.~E., {Carley}, E.~P., {Hayes}, L.~A., {et~al.} 2019, Nature
  Astronomy, 3, 452, \dodoi{10.1038/s41550-019-0689-z}

\bibitem[{{Newkirk}(1961)}]{Newkirk1961}
{Newkirk}, Jr., G. 1961, \apj, 133, 983, \dodoi{10.1086/147104}

\bibitem[{{Nitta} {et~al.}(2013){Nitta}, {Schrijver}, {Title}, \&
  {Liu}}]{Nitta2013}
{Nitta}, N.~V., {Schrijver}, C.~J., {Title}, A.~M., \& {Liu}, W. 2013, \apj,
  776, 58, \dodoi{10.1088/0004-637X/776/1/58}

\bibitem[{{Ontiveros} \& {Vourlidas}(2009)}]{Ontiveros2009}
{Ontiveros}, V., \& {Vourlidas}, A. 2009, \apj, 693, 267,
  \dodoi{10.1088/0004-637X/693/1/267}

\bibitem[{{Pesnell} {et~al.}(2012){Pesnell}, {Thompson}, \&
  {Chamberlin}}]{Pesnell2012}
{Pesnell}, W.~D., {Thompson}, B.~J., \& {Chamberlin}, P.~C. 2012, \solphys,
  275, 3, \dodoi{10.1007/s11207-011-9841-3}

\bibitem[{{Priest}(2014)}]{Priest2014}
{Priest}, E. 2014, {Magnetohydrodynamics of the Sun}

\bibitem[{{Qin} {et~al.}(2018){Qin}, {Kong}, \& {Zhang}}]{Qin2018}
{Qin}, G., {Kong}, F.~J., \& {Zhang}, L.~H. 2018, \apj, 860, 3,
  \dodoi{10.3847/1538-4357/aac26f}

\bibitem[{{Reiner} {et~al.}(2003){Reiner}, {Vourlidas}, {Cyr}, {Burkepile},
  {Howard}, {Kaiser}, {Prestage}, \& {Bougeret}}]{Reiner2003}
{Reiner}, M.~J., {Vourlidas}, A., {Cyr}, O.~C.~S., {et~al.} 2003, \apj, 590,
  533, \dodoi{10.1086/374917}

\bibitem[{{Retino} {et~al.}(2019){Retino}, {Khotyaintsev}, {Le Contel},
  {Marcucci}, {Plaschke}, {Vaivads}, {Angelopoulos}, {Blasi}, {De Keyser},
  {Dunlop}, {Dai}, {Eastwood}, {Fu}, {Haaland}, {Hoshino}, {Johlander},
  {Kepko}, {Kucharek}, {Lapenta}, {Lavraud}, {Maland raki}, {Matthaeus},
  {McWilliams}, {Petrukovich}, {Pin{\c{c}}on}, {Saito}, {Sorriso-Valvo},
  {Vainio}, \& {Wimmer-Schweingruber}}]{Retino2019}
{Retino}, A., {Khotyaintsev}, Y., {Le Contel}, O., {et~al.} 2019, arXiv
  e-prints, arXiv:1909.02783

\bibitem[{{Reynoso} {et~al.}(2013){Reynoso}, {Hughes}, \&
  {Moffett}}]{Reynoso2013}
{Reynoso}, E.~M., {Hughes}, J.~P., \& {Moffett}, D.~A. 2013, \aj, 145, 104,
  \dodoi{10.1088/0004-6256/145/4/104}

\bibitem[{{Rouillard} {et~al.}(2016){Rouillard}, {Plotnikov}, {Pinto},
  {Tirole}, {Lavarra}, {Zucca}, {Vainio}, {Tylka}, {Vourlidas}, {De Rosa},
  {Linker}, {Warmuth}, {Mann}, {Cohen}, \& {Mewaldt}}]{Rouillard2016}
{Rouillard}, A.~P., {Plotnikov}, I., {Pinto}, R.~F., {et~al.} 2016, \apj, 833,
  45, \dodoi{10.3847/1538-4357/833/1/45}

\bibitem[{{Ruan} {et~al.}(2018){Ruan}, {Yan}, {He}, {Zhang}, {Wang}, \&
  {Wei}}]{Ruan2018}
{Ruan}, W., {Yan}, L., {He}, J., {et~al.} 2018, \apj, 860, 99,
  \dodoi{10.3847/1538-4357/aac0f8}

\bibitem[{{Schatten} {et~al.}(1969){Schatten}, {Wilcox}, \&
  {Ness}}]{Schatten1969}
{Schatten}, K.~H., {Wilcox}, J.~M., \& {Ness}, N.~F. 1969, \solphys, 6, 442,
  \dodoi{10.1007/BF00146478}

\bibitem[{{Schrijver} \& {De Rosa}(2003)}]{Schrijver2003}
{Schrijver}, C.~J., \& {De Rosa}, M.~L. 2003, \solphys, 212, 165,
  \dodoi{10.1023/A:1022908504100}

\bibitem[{{Smith} {et~al.}(2003){Smith}, {Khanzadyan}, \& {Davis}}]{Smith2003}
{Smith}, M.~D., {Khanzadyan}, T., \& {Davis}, C.~J. 2003, \mnras, 339, 524,
  \dodoi{10.1046/j.1365-8711.2003.06195.x}

\bibitem[{{Su} {et~al.}(2016){Su}, {Cheng}, {Ding}, {Chen}, {Ning}, \&
  {Ji}}]{Su2016}
{Su}, W., {Cheng}, X., {Ding}, M.~D., {et~al.} 2016, \apj, 830, 70,
  \dodoi{10.3847/0004-637X/830/2/70}

\bibitem[{{Su} {et~al.}(2015){Su}, {Cheng}, {Ding}, {Chen}, \& {Sun}}]{Su2015}
{Su}, W., {Cheng}, X., {Ding}, M.~D., {Chen}, P.~F., \& {Sun}, J.~Q. 2015,
  \apj, 804, 88, \dodoi{10.1088/0004-637X/804/2/88}

\bibitem[{{Su} {et~al.}(2018{\natexlab{a}}){Su}, {Guo}, {Erd{\'e}lyi}, {Ning},
  {Ding}, {Cheng}, \& {Tan}}]{Su2018}
{Su}, W., {Guo}, Y., {Erd{\'e}lyi}, R., {et~al.} 2018{\natexlab{a}}, Scientific
  Reports, 8, 4471, \dodoi{10.1038/s41598-018-22796-7}

\bibitem[{{Su} {et~al.}(2021){Su}, {Wang}, {Zhou}, {Lu}, {Zhou}, {Li}, {Shi},
  {Hu}, {Zhou}, {Wang}, {Yeh}, {Wang}, \& {Chen}}]{Su2021}
{Su}, W., {Wang}, Y., {Zhou}, C., {et~al.} 2021, \apj, 914, 139,
  \dodoi{10.3847/1538-4357/abfc49}

\bibitem[{{Su} {et~al.}(2018{\natexlab{b}}){Su}, {Veronig}, {Hannah}, {Cheung},
  {Dennis}, {Holman}, {Gan}, \& {Li}}]{SuY2018}
{Su}, Y., {Veronig}, A.~M., {Hannah}, I.~G., {et~al.} 2018{\natexlab{b}},
  \apjl, 856, L17, \dodoi{10.3847/2041-8213/aab436}

\bibitem[{{SunPy Community} {et~al.}(2015){SunPy Community}, {Mumford},
  {Christe}, {P{\'e}rez-Su{\'a}rez}, {Ireland}, {Shih}, {Inglis}, {Liedtke},
  {Hewett}, {Mayer}, {Hughitt}, {Freij}, {Meszaros}, {Bennett}, {Malocha},
  {Evans}, {Agrawal}, {Leonard}, {Robitaille}, {Mampaey}, {Iv{\'a}n
  Campos-Rozo}, \& {Kirk}}]{SunPy2015}
{SunPy Community}, T., {Mumford}, S.~J., {Christe}, S., {et~al.} 2015,
  Computational Science and Discovery, 8, 014009,
  \dodoi{10.1088/1749-4699/8/1/014009}

\bibitem[{{Susino} {et~al.}(2015){Susino}, {Bemporad}, \&
  {Mancuso}}]{Susino2015}
{Susino}, R., {Bemporad}, A., \& {Mancuso}, S. 2015, \apj, 812, 119,
  \dodoi{10.1088/0004-637X/812/2/119}

\bibitem[{{Vourlidas} {et~al.}(2013){Vourlidas}, {Lynch}, {Howard}, \&
  {Li}}]{Vourlidas2013}
{Vourlidas}, A., {Lynch}, B.~J., {Howard}, R.~A., \& {Li}, Y. 2013, \solphys,
  284, 179, \dodoi{10.1007/s11207-012-0084-8}

\bibitem[{{Vr{\v s}nak} \& {Cliver}(2008)}]{Vrsnak2008}
{Vr{\v s}nak}, B., \& {Cliver}, E.~W. 2008, \solphys, 253, 215,
  \dodoi{10.1007/s11207-008-9241-5}

\bibitem[{{Vr{\v s}nak} {et~al.}(2002){Vr{\v s}nak}, {Magdaleni{\'c}},
  {Aurass}, \& {Mann}}]{Vrsnak2002}
{Vr{\v s}nak}, B., {Magdaleni{\'c}}, J., {Aurass}, H., \& {Mann}, G. 2002,
  \aap, 396, 673, \dodoi{10.1051/0004-6361:20021413}

\bibitem[{{Wang} {et~al.}(2018){Wang}, {Shen}, {Liu}, {Liu}, {Guo}, {Li}, {Xu},
  {Hu}, \& {Zhang}}]{Wang2018}
{Wang}, Y., {Shen}, C., {Liu}, R., {et~al.} 2018, Journal of Geophysical
  Research (Space Physics), 123, 3238, \dodoi{10.1002/2017JA024971}

\bibitem[{{Wang} \& {Sheeley}(1992)}]{Wang1992}
{Wang}, Y.-M., \& {Sheeley}, Jr., N.~R. 1992, \apj, 392, 310,
  \dodoi{10.1086/171430}

\bibitem[{{Weber} {et~al.}(2004){Weber}, {Deluca}, {Golub}, \&
  {Sette}}]{Weber2004}
{Weber}, M.~A., {Deluca}, E.~E., {Golub}, L., \& {Sette}, A.~L. 2004, in IAU
  Symposium, Vol. 223, Multi-Wavelength Investigations of Solar Activity, ed.
  A.~V. {Stepanov}, E.~E. {Benevolenskaya}, \& A.~G. {Kosovichev}, 321--328,
  \dodoi{10.1017/S1743921304006088}

\bibitem[{{Wild}(1950)}]{Wild1950}
{Wild}, J.~P. 1950, Australian Journal of Scientific Research A Physical
  Sciences, 3, 541

\bibitem[{{Wu} {et~al.}(1986){Wu}, {Steinolfson}, \& {Zhou}}]{Wu1986}
{Wu}, C.~S., {Steinolfson}, R.~S., \& {Zhou}, G.~C. 1986, \apj, 309, 392,
  \dodoi{10.1086/164611}

\bibitem[{{Wuelser} {et~al.}(2004){Wuelser}, {Lemen}, {Tarbell}, {Wolfson},
  {Cannon}, {Carpenter}, {Duncan}, {Gradwohl}, {Meyer}, {Moore}, {Navarro},
  {Pearson}, {Rossi}, {Springer}, {Howard}, {Moses}, {Newmark},
  {Delaboudiniere}, {Artzner}, {Auchere}, {Bougnet}, {Bouyries}, {Bridou},
  {Clotaire}, {Colas}, {Delmotte}, {Jerome}, {Lamare}, {Mercier}, {Mullot},
  {Ravet}, {Song}, {Bothmer}, \& {Deutsch}}]{Wuelser2004}
{Wuelser}, J.-P., {Lemen}, J.~R., {Tarbell}, T.~D., {et~al.} 2004, in Procspie
  SPIE, Vol. 5171, Telescopes and Instrumentation for Solar Astrophysics, ed.
  S.~{Fineschi} \& M.~A. {Gummin}, 111--122, \dodoi{10.1117/12.506877}

\bibitem[{{Ying} {et~al.}(2019){Ying}, {Bemporad}, {Giordano}, {Pagano},
  {Feng}, {Lu}, {Li}, \& {Gan}}]{Ying2019}
{Ying}, B., {Bemporad}, A., {Giordano}, S., {et~al.} 2019, \apj, 880, 41,
  \dodoi{10.3847/1538-4357/ab2713}

\bibitem[{{Zhang} {et~al.}(2021){Zhang}, {Wang}, \& {Kontar}}]{Zhang2021}
{Zhang}, P., {Wang}, C., \& {Kontar}, E.~P. 2021, \apj, 909, 195,
  \dodoi{10.3847/1538-4357/abd8c5}

\bibitem[{{Zhao} {et~al.}(2014){Zhao}, {Chen}, \& {Wu}}]{Zhao2014}
{Zhao}, G.~Q., {Chen}, L., \& {Wu}, D.~J. 2014, \apj, 786, 47,
  \dodoi{10.1088/0004-637X/786/1/47}

\bibitem[{{Zheleznyakov}(1970)}]{Zheleznyakov1970}
{Zheleznyakov}, V.~V. 1970, {Radio emission of the sun and planets}

\bibitem[{{Zucca} {et~al.}(2014{\natexlab{a}}){Zucca}, {Carley}, {Bloomfield},
  \& {Gallagher}}]{Zucca2014a}
{Zucca}, P., {Carley}, E.~P., {Bloomfield}, D.~S., \& {Gallagher}, P.~T.
  2014{\natexlab{a}}, \aap, 564, A47, \dodoi{10.1051/0004-6361/201322650}

\bibitem[{{Zucca} {et~al.}(2014{\natexlab{b}}){Zucca}, {Pick}, {D{\'e}moulin},
  {Kerdraon}, {Lecacheux}, \& {Gallagher}}]{Zucca2014b}
{Zucca}, P., {Pick}, M., {D{\'e}moulin}, P., {et~al.} 2014{\natexlab{b}}, \apj,
  795, 68, \dodoi{10.1088/0004-637X/795/1/68}

\bibitem[{{Zucca} {et~al.}(2018){Zucca}, {Morosan}, {Rouillard}, {Fallows},
  {Gallagher}, {Magdalenic}, {Klein}, {Mann}, {Vocks}, {Carley}, {Bisi},
  {Kontar}, {Rothkaehl}, {Dabrowski}, {Krankowski}, {Anderson}, {Asgekar},
  {Bell}, {Bentum}, {Best}, {Blaauw}, {Breitling}, {Broderick}, {Brouw},
  {Br{\"u}ggen}, {Butcher}, {Ciardi}, {de Geus}, {Deller}, {Duscha},
  {Eisl{\"o}ffel}, {Garrett}, {Grie{\ss}meier}, {Gunst}, {Heald}, {Hoeft},
  {H{\"o}randel}, {Iacobelli}, {Juette}, {Karastergiou}, {van Leeuwen},
  {McKay-Bukowski}, {Mulder}, {Munk}, {Nelles}, {Orru}, {Paas}, {Pandey},
  {Pekal}, {Pizzo}, {Polatidis}, {Reich}, {Rowlinson}, {Schwarz}, {Shulevski},
  {Sluman}, {Smirnov}, {Sobey}, {Soida}, {Thoudam}, {Toribio}, {Vermeulen},
  {van Weeren}, {Wucknitz}, \& {Zarka}}]{Zucca2018}
{Zucca}, P., {Morosan}, D.~E., {Rouillard}, A.~P., {et~al.} 2018, \aap, 615,
  A89, \dodoi{10.1051/0004-6361/201732308}

\end{thebibliography}
\bibliographystyle{aasjournal}

\end{CJK*}
\end{document}